\documentstyle{article}

\catcode`\@=11
\def\marginnote#1{}

\newcount\hour
\newcount\minute
\newtoks\amorpm
\hour=\time\divide\hour by60
\minute=\time{\multiply\hour by60 \global\advance\minute by-\hour}
\edef\standardtime{{\ifnum\hour<12 \global\amorpm={am}%
        \else\global\amorpm={pm}\advance\hour by-12 \fi
        \ifnum\hour=0 \hour=12 \fi
        \number\hour:\ifnum\minute<10 0\fi\number\minute\the\amorpm}}
\edef\militarytime{\number\hour:\ifnum\minute<10 0\fi\number\minute}

\def\draftlabel#1{{\@bsphack\if@filesw {\let\thepage\relax
   \xdef\@gtempa{\write\@auxout{\string
      \newlabel{#1}{{\@currentlabel}{\thepage}}}}}\@gtempa
   \if@nobreak \ifvmode\nobreak\fi\fi\fi\@esphack}
        \gdef\@eqnlabel{#1}}
\def\@eqnlabel{}
\def\@vacuum{}
\def\draftmarginnote#1{\marginpar{\raggedright\scriptsize\tt#1}}

\def\draft{\oddsidemargin -.5truein
        \def\@oddfoot{\sl preliminary draft \hfil
        \rm\thepage\hfil\sl\today\quad\militarytime}
        \let\@evenfoot\@oddfoot \overfullrule 3pt
        \let\label=\draftlabel
        \let\marginnote=\draftmarginnote
   \def\@eqnnum{(\theequation)\rlap{\kern\marginparsep\tt\@eqnlabel}%
\global\let\@eqnlabel\@vacuum}  }
\def\bea{\begin{eqnarray}}
\def\eea{\end{eqnarray}}
\def\nn{\nonumber}

\def\beq{\begin{equation}}
\def\eeq{\end{equation}}
\def\ba{\beq\new\begin{array}{c}}
\def\ea{\end{array}\eeq}
\def\be{\ba}
\def\ee{\ea}
\def\stackreb#1#2{\mathrel{\mathop{#2}\limits_{#1}}}
\def\Tr{{\rm Tr}}
\def\res{{\rm res}}

\def\rank{{\rm rank}}
\def\2{{1\over 2}}
\def\N2{${\cal N}=2$}

\def\Bf#1{\mbox{\boldmath $#1$}}
\def\balpha{{\Bf\alpha}}

\def\bphi{{\Bf\phi}}

\def\bPhi{{\Bf\Phi}}

\def\bomega{{\bfit\omega}}

\def\d{\partial}

\begingroup\ifx\undefined\newsymbol \else\def\input#1 {\endgroup}\fi
\input amssym.def \relax
\input amssym
\newfont{\hr}{msbm10}
\newfont{\ams}{msam10}
\font\teneufm=cmmib10
\font\seveneufm=cmmib7
\font\fiveeufm=cmmib5
\def\bfit#1{{\textfont1=\teneufm\scriptfont1=\seveneufm
\scriptscriptfont1=\fiveeufm
\mathchoice{\hbox{$\displaystyle#1$}}{\hbox{$\textstyle#1$}}
{\hbox{$\scriptstyle#1$}}{\hbox{$\scriptscriptstyle#1$}}}}

\makeatletter
\newdimen\normalarrayskip              
\newdimen\minarrayskip                 
\normalarrayskip\baselineskip
\minarrayskip\jot
\newif\ifold             \oldtrue            \def\new{\oldfalse}
\def\arraymode{\ifold\relax\else\displaystyle\fi} 
\def\eqnumphantom{\phantom{(\theequation)}}     
\def\@arrayskip{\ifold\baselineskip\z@\lineskip\z@
     \else
     \baselineskip\minarrayskip\lineskip2\minarrayskip\fi}
\def\@arrayclassz{\ifcase \@lastchclass \@acolampacol \or
\@ampacol \or \or \or \@addamp \or
   \@acolampacol \or \@firstampfalse \@acol \fi
\edef\@preamble{\@preamble
  \ifcase \@chnum
     \hfil$\relax\arraymode\@sharp$\hfil
     \or $\relax\arraymode\@sharp$\hfil
     \or \hfil$\relax\arraymode\@sharp$\fi}}
\def\@array[#1]#2{\setbox\@arstrutbox=\hbox{\vrule
     height\arraystretch \ht\strutbox
     depth\arraystretch \dp\strutbox
     width\z@}\@mkpream{#2}\edef\@preamble{\halign
\noexpand\@halignto
\bgroup \tabskip\z@ \@arstrut \@preamble \tabskip\z@ \cr}%
\let\@startpbox\@@startpbox \let\@endpbox\@@endpbox
  \if #1t\vtop \else \if#1b\vbox \else \vcenter \fi\fi
  \bgroup \let\par\relax
  \let\@sharp##\let\protect\relax
  \@arrayskip\@preamble}
\def\eqnarray{\stepcounter{equation}%
              \let\@currentlabel=\theequation
              \global\@eqnswtrue
              \global\@eqcnt\z@
              \tabskip\@centering
              \let\\=\@eqncr
              $$%
 \halign to \displaywidth\bgroup
    \eqnumphantom\@eqnsel\hskip\@centering
    $\displaystyle \tabskip\z@ {##}$%
    \global\@eqcnt\@ne \hskip 2\arraycolsep
         $\displaystyle\arraymode{##}$\hfil
    \global\@eqcnt\tw@ \hskip 2\arraycolsep
         $\displaystyle\tabskip\z@{##}$\hfil
         \tabskip\@centering
    &{##}\tabskip\z@\cr}

\begin{document}


\begin{titlepage}
\setcounter{footnote}0
\begin{center}
\hfill FIAN/TD-17/96\\
\hfill ITEP/TH-52/96\\
\vspace{0.3in}
{\LARGE\bf On Integrable Systems and Supersymmetric Gauge Theories }
\\
\bigskip
\bigskip
\bigskip
{\Large A.Marshakov
\footnote{E-mail address: mars@lpi.ac.ru, andrei@heron.itep.ru,
marshakov@nbivms.nbi.dk}}
\\
\bigskip
{\it Theory Department,  P. N. Lebedev Physics
Institute , Leninsky prospect, 53, Moscow,~117924, Russia\\
and ITEP, Moscow 117259, Russia}
\end{center}
\bigskip \bigskip

\begin{abstract}
The properties of the ${\cal N}=2$ SUSY gauge theories underlying the
Seiberg-Witten hypothesis are discussed. The main ingredients of the
formulation of the finite-gap solutions to integrable equations in terms of
complex curves and generating 1-differential are presented, the invariant
sense of these definitions is illustrated.
Recently found exact nonperturbative solutions to ${\cal N}=2$
SUSY gauge theories are formulated using the methods of the theory of
integrable systems and where possible the parallels between standard quantum
field theory results and solutions to integrable systems are discussed.
\end{abstract}

\end{titlepage}

\newpage
\setcounter{footnote}0

\section{Introduction: Main Definitions}

The aim of this paper is to present in a clear form the main ideas
of the relations between the exact solutions to the \N2 supersymmetric (SUSY)
Yang-\-Mills theory (arising in the point-particle limit of string theory) and
integrable systems.
The approach to the Seiberg-Witten effective theory based
on integrable systems was proposed in \cite{GKMMM} and developed along these
lines in \cite{MartW1}-\cite{Nekrasov}, where all necessary details can be
found.

The plan of the paper looks as follows. First, I review what is known about
construction of the effective actions for the low-energy \N2 Yang-Mills
theories and consider the definition
of the algebro-geometric solution to an integrable system in terms of a
complex curve and generating 1-differential. These definitions are
illustrated the basic example of the Toda
chain periodic solutions which is considered in detail. Next,
I pass to the Seiberg-Witten solutions directly and show that they
are indeed defined by the same data as the finite-gap solutions to
integrable systems, though the complete formulation requires to introduce
{\it deformations} of the finite-gap solutions. Finally, the explicit
differential equations and direct computations of the prepotential of the
effective theory are presented and compared when possible with the well-known
computations from supersymmetric quantum gauge theories.

\subsection{Seiberg-Witten effective theory}

Let us start with some introductory motivations and present the main
definitions which will be used below for the formulation of the effective
Seiberg-Witten theory in terms of integrable systems.

The object of study is given by the effective (abelian) \N2 supersymmetric
gauge theories in four (or five-)
\footnote{Let us notice immediately that the effective formulation of the
exact nonperturbative solutions we discuss here does not depend explicitely
on the dimension of target-space. Integrable systems suggest a {\em universal}
effective formulation which is known at present for some 2D string models as
well as for the theories considered in this paper.}
dimensions corresponding to the \N2 SUSY
Yang-Mills
theory with the bare action
\be\label{ym}
{\cal L} =\int d^4\theta \hat{\cal L}({\bPhi}) = \dots
 {1\over g^2}\Tr {\bf F}_{\mu\nu}^2 + i\theta \Tr {\bf F}_{\mu\nu}
\tilde{\bf F}_{\mu\nu} + \dots
\ee
The exact nonperturbative results
\cite{SW1,SW2,sun,fumat} contain
information about the spectrum of massive BPS excitations ("W-bosons" and
monopoles)
\footnote{The BPS ($\equiv$ the Bogomolny-\-Prasad-\-Sommerfeld) states are
the states in the "small" multiplet with masses being proportional to the
central charges of the extended ${\cal N}\geq 2$ algebra of supersymmetry.}
and the {\em Wilsonian} effective action for the massless
particles (see for example \cite{Wils}).
The most essential feature of this formulation is that the effective action
can be written in terms of a single (holomorphic) {\it function} of several
complex variables \cite{SW1,SW2}. Later on, according to the standard
terminology, this function will be referred to as {\em prepotential}.

For the \N2 SUSY gauge theory the result can be
understood in the following way. The scalar potential in the \N2 SUSY
gauge theory action has the form $V(\bphi ) =
\Tr [\bphi, \bphi ^{\dagger}]^2$ and its minima after factorization over the
gauge group correspond to the diagonal ($[\bphi, \bphi ^{\dagger}] = 0$),
in the theory with $SU(N_c)$ gauge group traceless matrices
\be\label{vacmatr}
\bphi = \left.\left(
\begin{array}{cccc}
A_1 &     &  &  \\
    & A_2 &  &  \\
    &     &\dots & \\
    &     &      & A_{N_c}
\end{array}\right)\right|_{\Tr\bphi = \sum A_j = 0}
\ee
whose invariants
\be\label{polyn}
\det (\lambda - \bphi) = P_{N_c}(\lambda ) = \sum_{k=0}^{N_c}
s_{N_c-k}\lambda ^k
\ee
(the total number of algebraically independent ones is
$\rank\ SU(N_c) = N_c -1$; one can also
take any other set of invariants,)
parameterize the moduli space of the theory. Due to the Higgs effect
the off-diagonal part of the gauge field
${\bf A}_{\mu}$ becomes massive, since
\be
[\bphi , {\bf A}_{\mu}]_{ij} = (A_i-A_j){\bf A}_{\mu}^{ij}
\ee
while the diagonal part, as it follows from (\ref{comm}) remains massless, i.e.
the gauge group $G = SU(N_c)$ breaks down to $U(1)^{\rank G} = U(1) ^{N_c -1}$.
Thus, the effective \N2 abelian gauge theory arises
with the effective Lagrangian, written in terms of the superfields
\be\label{sflds}
\Phi_i = \varphi^i + \vartheta\sigma_{\mu\nu}\tilde\vartheta f_{\mu\nu}^i +
\ldots
\nn \\
\sigma _{\mu\nu} \sim \left[\gamma _{\mu},\gamma _{\nu}\right]
\ee
whose vacuum values coincide with the diagonal values of (\ref{vacmatr}).
Therefore the function of complex variables ${\cal F}(a)
= \left.{\cal F}(A)\right|_{\sum A_i = 0}$, indeed determines the Wilsonian
effective action for the massless particles by means of the following
substitution
\be\label{subst}
{\cal L}_{\rm eff} \sim {\rm Im}\int d^4\vartheta
{\cal F}(A_i \rightarrow \Phi _i) =  \dots
{\rm Im}{\partial ^2{\cal F}\over\partial a_i\partial a_j}(a)f_{\mu\nu}^if_{\mu\nu}^j
+ \dots
\ee
This fact can be checked by explicit computations of quantum corrections
in conventional \N2 SUSY gauge theory.

As for the massive excitations, it turns out \cite{SW1,SW2}, that at least the
BPS massive spectrum, i.e. the spectrum of states of "small" multiplet
whose masses are proportional to the central charges of the extended
${\cal N}\geq 2$ SUSY algebra, is related with the prepotential ${\cal F}$
by $M \sim |{\bf na} + {\bf ma}_D|$,
where ${\bf a}_D = {\partial {\cal F}\over\partial {\bf a}}$. According
to the Seiberg-Witten hypothesis (which will be formulated strictly below
in subsect.1.3) the BPS masses ${\bf a}$ and ${\bf a}_D$ can be expressed
through the periods of a meromorphic differential on auxiliary
Riemann surface and depend on the vacuum expectation values of scalar
fields, which can be considered as certain co-ordinates on the moduli
space of auxiliary surface. For example, in the case of pure SUSY gauge
theory with the $SU(N_c)$ gauge group the auxiliary curve and meromorphic
differential have the form \cite{SW1,sun}
\be\label{suncu}
w + {1\over w} = 2P_{N_c}(\lambda)\ \ \ \ dS = \lambda{dw\over w}
\ee
while for the \N2 SUSY QCD \cite{SW2,fumat}
\be\label{sqcd}
W + {1\over W} = {2P_{N_c}(\lambda)\over P_{N_f}(\lambda)}
\ \ \ \ dS = \lambda{dW\over W}
\ee
Thus the knowledge of the function ${\cal F}$ and its derivatives as functions
on moduli (and also as functions of possible external sources {\bf T})
gives the most complete up to now information about the theory.
It will be demonstrated below that the fact that the nonperturbative solution
to the \N2 SUSY gauge
theory can be presented in terms of effective integrable system gives
rise to the main property of the ${\cal F}$-function, that it depends on
(part of) its variables as on (some) moduli of the complex structures of
(auxiliary) complex curves or Riemann surfaces (\ref{suncu}), (\ref{sqcd}).

\subsection{Integrable systems}

The main idea is to identify function
${\cal F}$ and other characteristics of effective theory with the objects
from the theory of integrable systems. Fortunately, it turns out that
this particular class of effective theories (as well as the class of
low-dimensional string models) can be described in terms of
the integrable systems of (Kadomtsev-Petviashvili) KP and Toda type.
The starting point is that the KP
\be\label{kadpet}
{\partial ^2U\over\partial T_2^{\ 2}} = {\partial\over\partial T_1}
\left( {\partial U\over\partial T_3} + U{\partial U\over\partial T_1} +
{\partial ^3U\over\partial T_1^{\ 3}}\right)
\ee
and the Toda lattice
\be\label{TodaLat}
{\partial ^2\phi _n\over\partial T_1\partial {\overline T}_1} =
e^{\phi_{n+1} -\phi_n}- e^{\phi_n-\phi_{n-1}}
\ee
equations (and other equations of the same class: to be referred to later as
KP/Toda type equations)
possess the infinite amount of integrals of motion which can
be considered as generators of the (mutually commuting and of course commuting
with the first flows (\ref{kadpet}) and (\ref{TodaLat})) infinite amount
(hierarchy) of flows parameterized by "elder" times. The differential
equations describing $T_k$-flows have a complicated form if written
for the functions (potentials) $U({\bf T})$ and $\phi _n({\bf T})$,
but there exists much more elegant way of presenting the whole picture.

This way is based on using the auxiliary linear problem for the hierarchy of
nonlinear equations
\be\label{auxhi}
{\partial\over\partial T_k}\Psi = B_k\Psi
\ee
where $B_k=B_k[U]$ (or $B_k=B_k[\phi ]$) are the differential operators
{\it only} in $T_1$ for the KP case (\ref{kadpet}) or difference operators
(with
respect to the discrete time $n$) in the case of Toda lattice
(\ref{TodaLat}). The solution $\Psi $ to the auxiliary linear problem
is called usually the Baker-Akhiezer (BA) function if the equations
(\ref{auxhi}) are supplemented by the Lax equation
\be\label{Lax}
{\cal L}\Psi = \lambda\Psi
\ee
which in the case of reductions of the KP/Toda hierarchies can be considered
as one of the equations of the tower (\ref{auxhi}). In such language the
hierarchy of nonlinear equations consisting of a "tower" built above
(\ref{kadpet}), (\ref{TodaLat}) becomes equivalent to a single (operator)
Lax equation
\be\label{Laxeq}
{\partial{\cal L}\over\partial T_k} = \left[ B_k, {\cal L}\right]
\ee
or to the consistency conditions (the Zakharov-Shabat equations)
\be\label{zasha}
\left[{\partial\over\partial T_k} - B_k,
{\partial\over\partial T_l} - B_l\right] = 0
\ee
The most universal object in such formulation is given by so called
Hirota's $\tau $-\-function, satisfying the infinite chain of the bilinear
equations (the Hirota equations) and generating solutions to the hierarchy,
the BA functions etc. For example, in the case of KP hierarchy the BA function
$\Psi (\lambda ,{\bf T})$ and the "potential" $U({\bf T})$ are expressed
through the $\tau $-function in the following way (in order to avoid
misleading below $\tau $-functions will be denoted as ${\cal T}$ while
notation $\tau$ is reserved for the modular parameter of elliptic curve):
\be\label{kphisol}
\Psi = e^{\sum T_k\lambda ^k}{{\cal T} \left(T_k - {1\over k\lambda ^k}\right)
\over{\cal T}({\bf T})}
\nn \\
U({\bf T}) = \partial ^2\log{\cal T} ({\bf T}) \equiv {\partial ^2\over
\partial T_1^2}\log{\cal T}({\bf T})
\nn \\
\dots
\ee
and one can easily find analogous formulas for the Toda lattice (see
for example \cite{UT84}) and other integrable systems.

The KP/Toda systems have infinite amount of solutions parameterized by
so called infinite-dimensional Grassmannian (roughly speaking a function
of two variables -- the initial conditions) \cite{Sato,SW85}. The particular
solutions can be distinguished by additional (sometimes linear) equations,
say, on the $\tau $-function.

A special role is played by {\it finite-dimensional} solutions to the
hierarchies of integrable equations when only the finite number of the
integrals of motion (and flows ${\d\over\d T_k}$) is algebraically independent.
The important example of such finite-dimensional integrable systems is
given by the finite-gap solutions, usually fixed by the Novikov
constraint \cite{TS}
\be\label{fingap}
\left[ {\cal L}, {\cal A}\right] = 0
\nn \\
{\cal A} = \sum _k^{\rm finite}c_kB_k
\ee
where ${\cal L}$ is the Lax operator (\ref{Lax}), $B_k$ are the evolution
operators (\ref{auxhi}), and $c_k$ -- some {\it finite} set of nonzero
constants. The integration of a generic finite-gap problem is given by the
Krichever construction \cite{Krico} and consists of the following steps
\footnote{Here only a "rough picture" of the Krichever construction is
presented, the exact mathematical definitions and theorems can be found in
\cite{Krico,TS,Dub,DKN,SW85}.}:
\begin{itemize}
\item The common spectrum of the commutative operators ${\cal L}$ and
${\cal A}$ (\ref{fingap}) can be described by a system of equations, giving
rise to a complex algebraic curve $\Sigma $.
\item The BA function is a section of some bundle over $\Sigma$ --
in our case it will be almost always a {\em line} bundle.
\item The moduli of a complex curve are the integrals of motion of the
system (\ref{fingap}).
\item The integrable change of variables is given by the Abel map, and
the Liouville torus (angle variables) is a real section of the Jacobian
of $\Sigma$.
\item The Hamiltonian structure of the finite-gap solution can be formulated
with the help of generating (meromorphic on $\Sigma$) 1-differential
$dS$, whose periods (the integrals over nontrivial cycles on Riemann
surface) are canonically normalized integrals of motion of given
dynamical system.
\end{itemize}
The arising complex curves are usually described by the algebraic equations
\be\label{eqcur}
{\cal P}(\lambda , w) = 0
\ee
(one relation on two complex variables (\ref{eqcur}), where ${\cal P}$ --
a polynomial, whose coefficients are moduli of the complex structure, defines
a $1_{\bf C}d$ complex or $2_{\bf R}d$ real manifold) or by system of equations on several
complex variables. Topologically any complex curve is characterized by
a single non-negative integer parameter -- genus $g$ (the number of handles),
and for a given genus the complex structure is parameterized by
$3g-3$ complex numbers -- the moduli of the complex structure
$\dim _{\bf C}{\cal M}_g = 3g-3$.

The finite gap integrable systems usually correspond to the
$g$-parametric families of complex curves (so that the dimension of the
corresponding submanifold in the moduli space is equal to the dimension
of the Jacobian, i.e. the number of independent integrals of motion
coincides with the number of angle variables
\footnote{In fact the $g$-parametric families arise in the simplest cases
(in the context of this paper these are the theories with the $SU(N_c)$ gauge
group). In general situation one should consider the Prym manifold arising
as a "factor" of Jacobian over some involution (see for explicit examples
\cite{MMM3} and references therein).
}). The dimension of the Jacobian
is determined by the total number of (globally defined) holomorphic
differentials
$d\omega _i$, $i = 1,\dots,g$, and is equal to the genus of $\Sigma _g$.
On $\Sigma _g$ there exists $2g$ independent noncontractable contours
(two around each handle) which can be canonically split into so called
$A_i$, $i = 1,\dots,g$, and $B_i$, $i = 1,\dots,g$ cycles with the
intersection index $A_i\circ B_j = \delta _{ij}$. The holomorphic differentials
are usually taken to be normalized to the ${\bf A}$-cycles
\be\label{normA}
\oint _{A_j}d\omega _i = \delta _{ij}
\ee
then the integrals of the ${\bf B}$-cycles give the period matrix
\be\label{pemat}
\oint _{B_j}d\omega _i =  T_{ij}
\nn \\
\int _{\Sigma _g}d\omega _i\wedge \overline {d\omega}_j = {\rm Im} T_{ij}
\ee
As it is well known the period matrix (\ref{pemat}) is symmetric which can be checked
by application of the Stokes theorem to
\be\label{symmpemat}
0 = \int _{\Sigma _g}d\omega _i\wedge d\omega _j =
\sum _{k = 1}^g \oint _{A_k}d\omega _i\oint _{B_k}d\omega _j -
(i\leftrightarrow j) = T_{ij} - T_{ji}
\ee
The derivatives of generating differential $dS$ over $g$ directions in moduli
space, corresponding
to the integrals of motion give rise to (some) holomorphic differentials
\be\label{hol}
{\partial dS\over\partial h_k}\sim dv_k
\ee
where in fact the canonical holomorphic differential appear only if one takes
as co-ordinates on moduli space the
canonically normalized integrals of motion -- ${\bf A}$-periods of the
differential $dS$
\be\label{aper}
{\bf a} = \oint _{\bf A} dS
\ee
By accepted convention the corresponding "dual" ${\bf B}$-periods will be
called ${\bf a}_D$
\be\label{adper}
{\bf a}_D = \oint _{\bf B}dS
\ee
The existence of the relation (\ref{hol})  can be trivially checked for all
known $g$-parametric families of curves. In fact it is related to the
existence of specific co-ordinates on moduli space satisfying the
consistency condition
\be\label{consist}
{\partial dv_k\over\partial h_l} =
{\partial dv_l\over\partial h_k}
\ee
This property will be discussed in more general context of the Whitham
hierarchy.
The finite-gap solutions are the most simple examples of solutions to
integrable equations which are related directly to the nonperturbative
quantum theories. In fact they should be considered only as some
"approximations" to the exact solutions, which allow one however to
to describe part of the {\em exact} physical characteristics of the
theory, mostly concerning its massless sector. Moreover, in many cases
(e.g. in 2D string theories) the exact solutions can be considered as
integrable deformations of the finite-gap solutions, described
in terms of the Whitham hierarchies.

\subsection{Seiberg-Witten map}

Now we can pass to the exact formulation of the Seiberg-Witten effective theory
which can be formally defined as a map
\be\label{ymval}
G,\tau ,h_k \rightarrow a_i,\ a_i^D
\ee
($G$ is gauge group, $\tau$ -- the UV coupling
constant, $h_k = {1\over k}\langle\Tr\Phi ^k\rangle$ -- the v.e.v.'s of the
Higgs field)
and has an elegant description in terms of an integrable system. In most
known cases the integrable system is described in terms of a complex curve
$\Sigma _g$ with $h_k$ parameterizing some of the
(in most cases hyperelliptic) moduli of complex structures. The map (\ref{ymval})
is described in terms of the periods (\ref{aper}), (\ref{adper}) of the
meromorphic 1-form (\ref{hol}) which determine the BPS massive
spectrum,
\be\label{bps}
M \sim |{\bf na} + {\bf ma}^D|
\ee
and the prepotential ${\cal F}$ by the defining equation
\be\label{prepot}
a_i^D = {\partial{\cal F}\over\partial a_i}
\ee
The derivatives give the period matrix having the sense
of the low-energy coupling constants for the abelian gauge fields
(cf. with (\ref{subst})):
indeed, from (\ref{aper}), (\ref{adper}) and (\ref{pemat}) it follows that
\be\label{pematF}
\delta _{ij} ={\d a_i\over \d a_j} =  \oint _{A_i}{\d dS\over\d a_j} =
\oint _{A_i} d\omega _j
\nn \\
{\partial ^2{\cal F}\over \partial a_i\partial a_j} =
{\partial a^D_i\over\partial a_j} =
\oint _{B_i}{\d dS\over\d a_j}= \oint _{B_i} d\omega _j = T_{ij}
\ee
The curves $\Sigma _{g = \rank G}$ are special spectral curves
of the nontrivial finite-gap solutions to the periodic Toda-chain problem and
its natural deformations. The main object to be considered below -- the
prepotential
\be\label{flogt}
{\cal F} = \log{\cal T}
\ee
is logariphm of the $\tau $-function of the Whitham hierarchy, associated to
a particular finite gap solution.

\section{Toda chain: the periodic problem}

Let us demonstrate the above formulas on the simplest Toda-chain model, which
in the framework of the nonperturbative solutions
corresponds to the $4d$ {\it pure} gauge ${\cal N}=2$ supersymmetric
Yang-Mills theory (or \N2 SUSY gluodynamics) \cite{sun,GKMMM}. The periodic
problem in this model can be formulated
in two different ways, which could be further deformed into two different
directions. These deformations
are hypothetically related to the two different couplings of the $4d$
theory by adding the adjoint and fundamental matter
${\cal N}=2$ hypermultiplets correspondingly \cite{WiDo}-\cite{IM},
\cite{Spin,Spin2}.

The Toda chain system is a simple system of particles where only the neighbor
ones interact with the exponential potential and can be defined by the
equations of motion
\be\label{Todaeq}
\frac{\partial \phi _i}{\partial t} = p_i \ \ \ \ \
\frac{\partial p_i}{\partial t} = e^{\phi _{i+1} -\phi _i}- e^{\phi _i-\phi _{i-1}}
\ee
where one assumes (for the periodic problem with the ``period" $N_c$) that
$\phi _{i+N_c} = \phi _i$ and
$p_{i+N_c} = p_i$. It is an integrable system, with $N_c$
Poisson-commuting Hamiltonians, $h_1^{TC} = \sum p_i$, $h_2^{TC} =
\sum\left(\frac{1}{2}p_i^2 + e^{\phi _i-\phi _{i-1}}\right)$, etc. As
any finite-gap solution the periodic problem in Toda chain is formulated
in terms of (the eigenvalues and the eigenfunctions of) two operators:
the Lax operator (\ref{Lax}) ${\cal L}$ (or the auxiliary linear problem for
(\ref{Todaeq}))
\be\label{laxtoda}
\lambda\psi ^{\pm}_n =
\sum _k {\cal L}_{nk}\psi ^{\pm}_k =
e^{{1\over 2}(\phi _{n+1}-\phi _n)}\psi ^{\pm}_{n+1} + p_n\psi ^{\pm}_n +
e^{{1\over 2}(\phi _n-\phi _{n-1})}
\psi ^{\pm}_{n-1}
\ (=  \pm {\partial\over\partial t}\psi ^{\pm}_n)
\ee
and the second (${\cal A}$-operator (\ref{fingap})) in this case can be chosen
as a {\it monodromy} or shift operator in a discrete
variable -- the number of a particle
\be\label{T-op}
T\phi _n = \phi _{n+N_c}\ \ \ \ \ \ Tp_n = p_{n+N_c}\ \ \ \ \ \ \ \
T\psi_n = \psi_{n+N_c}
\ee
The common spectrum of these two operators
\footnote{Let us point out that we consider a {\it periodic} problem for
the Toda chain when only the BA function can acquire a nontrivial factor
under the action of the shift operator while the coordinates and momenta
themselves are periodic. The {\it quasi}periodicity of coordinates and momenta
-- when they acquire a nonzero shift -- corresponds to the change of
the coupling constant in the Toda chain Hamiltonians.}
\be\label{spec}
{\cal L}\psi = \lambda\psi \ \ \ \ \ T\psi = w\psi\ \ \ \ \ \ [{\cal L},T]=0
\ee
mean that there exists a relation between them ${\cal P}({\cal L},T) = 0$
which can be strictly formulated in terms of spectral curve $\Sigma$:
${\cal P}(\lambda,w) = 0$. The
generation function for these Hamiltonians can be written in terms of
${\cal L}$ and $T$ operators and the Toda chain possesses two
essentially different formulations of this kind.

In the first version the Lax operator (\ref{laxtoda}) is written in the
basis of the $T$-operator eigenfunctions and becomes the $N_c\times N_c$
matrix,
\be\label{LaxTC}
{\cal L}^{TC}(w) =
\left(\begin{array}{ccccc}
 p_1 & e^{{1\over 2}(\phi_2-\phi_1)} & 0 & & we^{{1\over 2}(\phi_1-\phi_{N_c})}\\
e^{{1\over 2}(\phi_2-\phi_1)} & p_2 & e^{{1\over 2}(\phi_3 - \phi_2)} & \ldots & 0\\
0 & e^{{1\over 2}(\phi_3-\phi_2)} & p_3 & & 0 \\
 & & \ldots & & \\
\frac{1}{w}e^{{1\over 2}(\phi_1-\phi_{N_c})} & 0 & 0 & & p_{N_c}
\end{array} \right)
\ee
defined on the two-punctured sphere. Matrix (\ref{LaxTC}) is almost
three-\-diagonal as it follows from (\ref{laxtoda}), the only extra nonzero
elements appear in the off-diagonal corners exactly due to periodic
conditions (\ref{T-op}) reducing therefore naively infinite-dimensional matrix
(\ref{laxtoda}) to a finite-dimensional one depending on the spectral parameter
$w$.
The eigenvalues of the Lax operator (\ref{LaxTC}) are defined from the spectral
equation
\be\label{SpeC}
{\cal P}(\lambda,w) = \det_{N_c\times N_c}\left({\cal L}^{TC}(w) -
\lambda\right) = 0
\ee
Substituting the explicit expression (\ref{LaxTC}) into (\ref{SpeC}),
one gets \cite{KriDu}:
\be\label{fsc-Toda}
w + \frac{1}{w} = 2P_{N_c}(\lambda )
\ee
or
\be\label{hypelTC}
y^2 = P_{N_c}^2(\lambda ) - 1 \ \ \ \ \ \ \ 2y = w - {1\over w}
\ee
where $P_{N_c}(\lambda )$ is a polynomial of degree $N_c$,
with the coefficients being the Schur polynomials of the Hamiltonians
$h_k = \sum_{i=1}^{N_c} p_i^k + \ldots$:
\be\label{Schura}
P_{N_c}(\lambda ) = \lambda ^{N_c} + h_1 \lambda ^{N_c-1} +
\frac{1}{2}(h_2-h_1^2)\lambda ^{N_c-2} + \ldots
\ee
The spectral equation depends only on the mutually Poisson-commuting
combinations of the dynamical variables --
the Hamiltonians or better action variables --
parameterizing (a subspace in the) moduli space of the complex structures of
the hyperelliptic curves $\Sigma^{TC}$ of genus $N_c - 1 = \rank SU(N_c)$.

An alternative description of the same system arises when one (before
imposing periodic conditions) {\it solves}
explicitly the auxiliary linear problem (\ref{laxtoda}) which is a
{\it second}-order
difference equation. To do it one just rewrites (\ref{laxtoda}) as
\be\label{recrel}
\psi _{i+1} = (\lambda - p_i)\psi _i - e^{\phi_i - {\phi_{i+1} + \phi_{i-1}\over 2}}
\psi_{i-1}
\ee
or, since the space of solutions is 2-dimensional
\footnote{The initial condition for the recursion relation (\ref{recrel})
consists of two arbitrary functions, say, $\psi _1$ and $\psi _2$.}
(denoted by $\psi ^{+}$
and $\psi ^{-}$ in (\ref{laxtoda})), it can be rewritten as
${\tilde\psi}_{i+1}
= L^{TC}_i(\lambda ){\tilde\psi}_i$ where $\tilde\psi _i$ is a set of
two-vectors and $L^{TC}_i$ -- a chain of $2\times 2$ Lax matrices.
After a simple "gauge" transformation these matrices can be written in the
form \cite{FT}
\be\label{LTC}
L^{TC}_i(\lambda) =
\left(\begin{array}{cc} p_i + \lambda & e^{\phi_i} \\ e^{-\phi_i} & 0
\end{array}\right), \ \ \ \ \ i = 1,\dots ,N_c
\ee
This form is convenient to check integrable properties using
the Hamiltonian language: the matrices (\ref{LTC}) obey {\it quadratic}
$r$-matrix Poisson bracket relations \cite{Skl} (equivalent to $\{ \phi_i,p_j\}
= \delta _{ij}$)
\be\label{quadrP} \left\{
L^{TC}_i(\lambda)\stackrel{\otimes}{,}L_j^{TC}(\lambda')\right\} =
\delta_{ij} \left[ r(\lambda - \lambda'), L^{TC}_i(\lambda)\otimes
L^{TC}_j(\lambda')\right]
\ee
with the ($i$-independent!) numerical rational
$r$-matrix $r(\lambda) = \frac{1}{\lambda } \sum_{a=1}^3 \sigma_a\otimes
\sigma^a$ satisfying the classical Yang-Baxter equation.
As a consequence, the transfer matrix
\be\label{monomat}
T_{N_c}(\lambda) =
\prod_{N_c \ge i\ge 1}^{\curvearrowleft} L_i(\lambda )
\ee
satisfies the same Poisson-\-bracket relation
\be
\left\{ T_{N_c}(\lambda)\stackrel{\otimes}{,}T_{N_c}(\lambda')\right\}
= \left[ r(\lambda - \lambda'), T_{N_c}(\lambda)\otimes
T_{N_c}(\lambda')\right]
\ee
and the integrals of motion of the Toda chain are generated
by another form of spectral equation
\be\label{specTC0}
\det_{2\times 2}\left( T^{TC}_{N_c}(\lambda )
- w\right) = w^2 - w\Tr T^{TC}_{N_c}(\lambda ) + \det T^{TC}_{N_c}(\lambda ) =
w^2 - w\Tr T^{TC}_{N_c}(\lambda ) + 1 = 0
\ee
or
\be\label{specTC}
{\cal P}(\lambda ,w) = \Tr T^{TC}_{N_c}(\lambda) - w - \frac{1}{w} =
2P_{N_c}(\lambda) - w - \frac{1}{w} = 0
\ee
(We used the fact that
$\det_{2\times 2} L^{TC}(\lambda) = 1$ leads to $\det_{2\times 2} T_{N_c}^{TC}
(\lambda) = 1$.) The r.h.s. of (\ref{specTC}) is a polynomial of degree $N_c$
in $\lambda$, with the coefficients being the integrals of motion since
\be\label{trcom}
\left\{ \Tr T_{N_c}(\lambda ), \Tr T_{N_c}(\lambda' )\right\} =
\Tr \left\{ T_{N_c}(\lambda )\stackrel{\otimes}{,}T_{N_c}(\lambda' )\right\} =
\nn \\
= \Tr \left[ r(\lambda - \lambda' ), T_{N_c}(\lambda )\otimes
T_{N_c}(\lambda' )\right] = 0
\ee
The generating differential (\ref{hol}) for the Toda chain has the form
\be\label{dS}
dS^{TC} = \lambda {dw\over w}
\ee
Indeed, its derivatives over $g$ moduli
-- the coefficients of the polynomial $P_{N_c}(\lambda )$ in
(\ref{Schura}), (\ref{specTC})
\be\label{deriv}
{\partial dS^{TC}\over\partial s_k} \equiv  \left.{\partial dS^{TC}\over
\partial s_k} \right| _{\lambda = const} =
\lambda d{\partial\over\partial s_k}\log w = 2\lambda d\left( {{\partial
P_{N_c}\over s_k}\over y}\right) \cong {\lambda ^{k + 1}d\lambda\over y}
\ee
are (up to total derivatives which is denoted by $\cong $) holomorphic
differentials on curve (\ref{fsc-Toda}), (\ref{hypelTC}), (\ref{specTC}).
One should consider the derivatives over moduli as taken for fixed
$\lambda $ -- see below more comments concerning this question. The
explicit formulas for the prepotentials of the integrable system considered
in this subsection will be presented in sect.3.

\subsection{Integrable deformations of the Toda chain: coupling to
matter \label{Cal}}

The $N_c\times N_c$ matrix Lax operator (\ref{LaxTC}) can be thought of as a
"degenerate" case of the Lax operator for the $GL(N_c)$ Calogero system
\cite{KriCal}
\be\label{LaxCal}
{\cal L}^{Cal}(\xi) =
\left({\bf pH} + \sum_{\balpha}F({\bf q\balpha}|\xi)
E_{\balpha}\right) = \nn \\
= \left(\begin{array}{cccc}
 p_1 & F(q_1-q_2|\xi) & \ldots &
F(q_1 - q_{N_c}|\xi)\\
F(q_2-q_1|\xi) & p_2 & \ldots &
F(q_2-q_{N_c}|\xi)\\
 & & \ldots  & \\
F(q_{N_c}-q_1|\xi) &
F(q_{N_c}-q_2|\xi)& \ldots &p_{N_c}
\end{array} \right)
\ee
The matrix elements $F(q|\xi) = m\frac{\sigma(q+\xi)}{\sigma(q)\sigma(\xi)}
e^{\zeta(q)\xi}$ are expressed in terms of the Weierstrass elliptic
functions and, thus, the Lax operator ${\cal L}(\xi)$ is
defined on the elliptic curve $E(\tau)$ (complex torus
with periods $\omega, \omega '$ and modulus $\tau = \frac{\omega '}{\omega}$).
The Calogero coupling constant is $m^2$,
where in the $4d$
interpretation $m$ plays the role of the mass of the adjoint matter
${\cal N}=2$ hypermultiplet breaking ${\cal N}=4$ SUSY down to ${\cal N}=2$
\cite{WiDo}.

From (\ref{LaxCal}) it follows that the spectral curve $\Sigma^{Cal}$ for the
$GL(N_c)$ Calogero system is given by:
\be\label{fscCal}
\det_{N_c\times N_c} \left({\cal L}^{Cal}(\xi) - \lambda\right) = 0
\ee
and is defined as coverning of the elliptic curve $E(\tau )$
\be\label{ell}
y^2 = (x -  e_1)(x - e_2)(x - e_3) \ \ \ \ \
\ee
with the canonical holomorphic 1-differential
\be
d\xi = 2\frac{dx}{y}
\ee
The BPS masses ${\bf a}$ and ${\bf a}_D$ are now the periods of the
generating 1-differential
\be\label{dSCal}
dS^{Cal} \cong \lambda d\xi
\ee
along the non-contractable contours on $\Sigma^{Cal}$
\footnote{Let us point out that
the curve (\ref{fscCal}) has genus $g = N_c$
(while in general the genus of the curve defined by $N_c\times N_c$ matrix
grows as $N_c^2$)
but the
integrable system is still $2(N_c - 1)$-dimensional, i.e. one of the periods
of (\ref{dSCal}) vanishes identically due to the symmetries of the curve
(\ref{fscCal}) and there are only $N_c - 1 = g - 1$ independent integrals of
motion. This is a particular case of the Prym manifold considered above
when one restricts himself only to those contours which are trivial after
projection to a "bare" curve.
}. Integrability of the Calogero-Moser model can be described for example
in terms of the following Poisson structure
\be\label{lin-r}
\left\{ {\cal L}(\xi)\stackrel{\otimes}{,}{\cal L}(\xi ')\right\}
= \left[ {\cal R}_{12}^{Cal}(\xi ,\xi '),\ {\cal L}(\xi)\otimes {\bf 1}\right]
-
\left[ {\cal R}_{21}^{Cal}(\xi ,\xi '),\ {\bf 1} \otimes {\cal L}(\xi') \right]
\ee
defined by {\it dynamical} elliptic ${\cal R}$-matrix \cite{Rmat},
which guarantees the involution of the eigenvalues of matrix ${\cal L}$.

In order to recover the Toda-chain system, one takes the double-scaling
limit \cite{Ino}, when $m$ and $-i\tau$ both go
to infinity and
\be
q_i-q_j={\2}\left[(i-j)\log m +(\phi_i-\phi_j)\right]
\ee
so that the dimensionless coupling $\tau$ gets
substituted by a dimensional parameter $\Lambda^{N_c} \sim
m^{N_c}e^{i\pi\tau}$. In this limit,
the elliptic curve $E(\tau)$ degenerates into the (two-punctured) Riemann
sphere with coordinate $w = e^{\xi }e^{i\pi\tau}$ so that
\be
dS^{Cal} \rightarrow dS^{TC} \cong \lambda\frac{dw}{w}
\ee
The Lax operator of the Calogero system turns into that of the
$N_c$-periodic Toda chain (\ref{LaxTC}):
\be
{\cal L}^{Cal}(\xi )d\xi \rightarrow {\cal L}^{TC}(w)\frac{dw}{w}
\ee
and the spectral curve acquires the form (\ref{SpeC}). That is why the
Calogero-Moser model can be considered as an elliptic deformation of the
Toda chain.
In contrast to the Toda case, (\ref{fscCal}) can {\it not} be rewritten in
the form (\ref{fsc-Toda}) and specific $w$-dependence of the spectral
equation (\ref{SpeC}) is not preserved by embedding of Toda into Calogero-Moser
particle
system. However, the form (\ref{fsc-Toda}) can be naturally preserved by
the alternative deformation of the Toda-chain system when it
is considered as (a particular case of) a spin-chain model.

In the simplest example of $N_c=2$,
the spectral curve $\Sigma^{Cal}$ has genus 2. Indeed,
in this particular case, eq.(\ref{fscCal}) turns into
\be\label{caln2}
{\cal P}(\lambda;x,y) = \lambda ^2 - h_2 + \frac{g^2}{\omega^2} x = 0
\ee
This equation says that with any value of $x$ one
associates two points of $\Sigma^{Cal}$
\be
\lambda = \pm\sqrt{h_2 - \frac{g^2}{\omega^2}x}
\ee
i.e.
it describes $\Sigma^{cal}$ as a double covering of the
elliptic curve $E(\tau)$ ramified at the points
$x = \left( {\omega\over g}\right)^2h_2$
and $x = \infty$. In fact, $x = \left( {\omega\over g}\right)^2h_2$
corresponds to a {\it pair} of
points on $E(\tau)$ distinguished by the sign of $y$. This would be true
for $x = \infty$
as well, but $x = \infty$ is one of the branch points in
our parameterization (\ref{ell}) of $E(\tau)$. Thus, the {\it two} cuts
between $x = \left( {\omega\over g}\right)^2h_2$ and $x=\infty$ on every
sheet of
$E(\tau)$ touching at the common end at $x=\infty$ become effectively
a {\it single} cut between $\left(\left( {\omega\over g}\right)^2h_2, +\right)$
and $\left(\left( {\omega\over g}\right)^2h_2, -\right)$. Therefore, we can
consider the spectral
curve $\Sigma^{Cal}$ as two tori $E(\tau)$ glued along one cut, i.e.
$\Sigma^{Cal}_{N_c=2}$ has genus 2.
It turns out to be a hyperelliptic curve (for $N_c = 2$ only!)
after substituting in (\ref{caln2}) $x$ from the second equation to the
first one.

Two holomorphic 1-differentials on $\Sigma^{Cal}$ ($g = N_c = 2$) can be
chosen to be
\be\label{holn2}
v = \frac{dx}{y} \sim \frac{\lambda d\lambda}{y}
\ \ \ \ \ \ \
V =  \frac{dx}{y\lambda}\sim \frac{d\lambda }{y}
\ee
so that
\be
dS \cong \lambda d\xi =
\sqrt{h_2 - \frac{g^2}{\omega^2}\wp(\xi )}d\xi =
\frac{dx}{y}\sqrt{h_2 - \frac{g^2}{\omega^2}x}
\ee
It is easy to check the basic property (\ref{hol}):
\be
\frac{\partial dS}{\partial h_2} \cong \frac{1}{2}\frac{dx}{y\lambda}
\ee
The fact that only one of two holomorphic 1-differentials (\ref{holn2})
appears at the r.h.s. is related to their different parity
with respect to the ${\bf Z}_2\otimes {\bf Z}_2$ symmetry of $\Sigma^{Cal}$:
$y \rightarrow -y$ and $\lambda \rightarrow -\lambda$.
Since $dS$ has certain parity, its
integrals along the two of four elementary non-contractable cycles
on $\Sigma^{Cal}$ automatically vanish leaving only two
non-vanishing quantities $a$ and $a_D$, as necessary for
the $4d$ interpretation. Moreover, two rest nonzero periods can be defined
in terms of the genus $1$ "reduced" curve
\be
Y^2 = (y\lambda)^2 = \left(h_2 - \frac{g^2}{\omega^2}x\right)
\prod_{a=1}^3 (x - e_a),
\ee
equiped with $dS \cong \left(h_2 - \frac{g^2}{\omega^2}x\right)\frac{dx}{Y}$.
Since for this curve $x = \infty$ is no more a ramification point, $dS$
has simple poles when $x = \infty$ (on both sheets of
$\Sigma^{Cal}_{reduced}$) with the residues $\pm\frac{g}{\omega} \sim \pm m$.

The opposite limit of the Calogero-Moser system with vanishing coupling
constant $g^2 \sim m^2 \rightarrow 0$ corresponds to the ${\cal N}=4$ SUSY
Yang-Mills theory with identically vanishing $\beta $-function.
The corresponding integrable
system is a collection of {\it free} particles and the generating differential
$dS \cong \sqrt{h_2}\cdot d\xi $ is just a {\it holomorphic} differential on
$E(\tau )$.

Now, let us turn to another deformation of the Toda chain corresponding to
the coupling of the ${\cal N}=2$ SYM theory to the fundamental matter.
According to \cite{SW2,fumat}, the spectral curves for
the ${\cal N}=2$ SQCD with any $N_f < 2N_c$ have the same
form as (\ref{spec}) with a less trivial monodromy matrix with the
invariants
\be\label{trdet}
\Tr\ T_{N_c}(\lambda) \equiv 2P_{N_c}(\lambda) = 2P^{(0)}_{N_c}(\lambda) +
R_{N_c-1}(\lambda),
\ \ \ \ \det T_{N_c}(\lambda ) = Q_{N_f}(\lambda ),
\ee
and $Q_{N_f}(\lambda)$ and $R_{N_c-1}(\lambda)$ are certain
$h$-{\it independent} polynomials of $\lambda$.

A natural proposal is to look at the orthogonal to the previous
generalization of the Toda chain,
i.e. deform eqs.(\ref{quadrP})-(\ref{trcom}) preserving the Poisson brackets
\be\label{quadr-r}
\left\{L(\lambda)\stackrel{\otimes}{,}L(\lambda')\right\} =
\left[ r(\lambda-\lambda'),\ L(\lambda)\otimes L(\lambda')\right],
\nn \\
\left\{T_{N_c}(\lambda)\stackrel{\otimes}{,}T_{N_c}(\lambda')\right\} =
\left[ r(\lambda-\lambda'),\
T_{N_c}(\lambda)\otimes T_{N_c}(\lambda')\right],
\ee
and, thus, the possibility to build a
monodromy matrix $T(\lambda)$ by multiplication of $L_i(\lambda)$'s.
The full spectral curve for the periodic spin chain is
still given by:
\be\label{fsc-SCh}
\det\left(T_{N_c}(\lambda) -  w\right) = 0,
\ee
but since in general $\det T_{N_c}(\lambda ) = \prod _{i = 1}^{N_c}\det
L(\lambda - \lambda _i) \neq 1$ equation (\ref{fsc-SCh}) acquires the form
\be\label{fsc-sc1}
w + \frac{\det T_{N_c}(\lambda)}{w} =
 \Tr T_{N_c}(\lambda ),
\ee
or
\be
W + \frac{1}{W} = \frac{\Tr T_{N_c}(\lambda)}
{\sqrt{\det T_{N_c}(\lambda)}} \equiv {2P_{N_c}(\lambda )\over\sqrt{
Q_{N_f}(\lambda )}}
\label{fsc-sc2}
\ee
while generating 1-form is now
\be\label{1f}
dS = \lambda\frac{dW}{W}
\ee
The r.h.s. of the equations (\ref{fsc-sc1}), (\ref{fsc-sc2}) contain the
dynamical variables of
the spin system only in the special combinations -- its
Hamiltonians (which are all in involution, i.e. Poisson-commuting).
The explicit examples can be found in \cite{Spin,Spin2}.

In this picture the rational $XXX$ spin chain literally corresponds to
a $N_f < 2N_c$ \N2 SUSY QCD while the conformal $N_f = 2N_c$ case when an
additional dimensionless parameter appears is hypothetically described by
the $XYZ$ chain with the Hamiltonian structure given by elliptic Sklyanin
algebra \cite{Skl} (see \cite{Spin2} for detailed analysis of this case
where however still a lot of open questions exist).

\subsection{Symplectic structure of the finite-gap systems}

Now let us turn to the discussion of a more subtle point -- why the
generating 1-form (\ref{hol}) indeed describes an integrable system.
To do this I consider, first, the simplest definition of the generating
1-form. According to this definition the generating form (\ref{hol})
defines the symplectic structure on the phase space of the finite-gap
solutions. This symplectic structure was introduced in
\cite{DKN} and recently proposed in \cite{KriPho} as coming directly
from the symplectic form on the space of all the solutions to the hierarchy.
Below I give the most simple and straightforward proof of this fact as
presented in \cite{M5,M7} which is supplied by concrete examples having
direct relation to the integrable systems arising in the formulation of
nonperturbative results in quantum theories. In addition we will discuss
the relation of generating differential to the duality transformation
in nonperturbative $c < 1$ string theory.

To prove that (\ref{dS}) is a generating one-form of the whole hierarchy
one starts with the variation of the generating function
\be\label{S}
S(\Sigma ,{\bf \gamma}) = \sum _i \int ^{\gamma _i}dS =
\sum _i \int ^{\gamma _i}Edp
\ee
(where $dE$ and $dp$ ($= d\lambda $ and $={dw\over w}$) in the particular case
above) are two meromorphic differentials (with fixed periods: for example
for the hyperelliptic coordinate $\lambda$ all $\oint d\lambda = 0$)
on a spectral curve $\Sigma $ and $\bf \gamma $ is the divisor
of the solution (i.e. the set of points -- with their multiplicities --
the poles of the BA function)). Taking the first variation
\footnote{The total variation, which includes the variation of the moduli
of complex structure of the curve is considered. To compare the differentials
on two curves with different complex structure one should introduce the
connection, which will be chosen below as satisfying the consition
$\delta _{moduli} p = 0$, i.e. such that function $p$ is covariantly constant
(one can compare it with \cite{KriPho}, where the role of covariantly
constant function is played by function $E$).},
one gets
\be\label{var}
\delta S = \sum _i (Edp)(\gamma _i) + \sum _i \int ^{\gamma _i}\delta E dp
\ \ \ \ \ \
\delta ^2 S = \delta\left( \sum _i (Edp)(\gamma _i)\right)
+ \sum _i (\delta E dp)(\gamma _i)
\ee
From $\delta ^2 S = 0$ it follows that
\be\label{wsform}
\varpi  = \delta E\wedge\delta p =
\delta\left( \sum _i (Edp)(\gamma _i)\right) =
- \sum _i (\delta E dp)(\gamma _i)
\ee
Now, the variation $\delta E$ (for constant $p$) follows from the Lax
equation (auxiliary linear problem)
\be\label{lax}
{\partial\over\partial t}\psi = {\cal L}\psi \ (= E\psi )
\ee
so that
\be\label{varE}
\delta E = {\langle \psi ^{\dagger}\delta {\cal L}\psi\rangle\over
\langle \psi ^{\dagger}\psi\rangle}
\ee
and one concludes that
\be\label{general}
\varpi =  - \langle \delta{\cal L}\sum _i \left( dp{\psi ^{\dagger}\psi\over
\langle \psi ^{\dagger}\psi\rangle}\right) (\gamma _i) \rangle
\ee
Let us turn to several important examples.\\
{\bf KP/KdV}. In the KP-case the equation (\ref{lax}) looks as
\be\label{laxkp}
{\partial\over\partial t}\psi = \left({\partial ^2\over\partial x^2} +
u\right) \psi \ (= E\psi )
\ee
therefore the equation (\ref{general}) implied by
$\langle \psi ^{\dagger}\psi\rangle = \int _{dx}\psi ^{\dagger}(x,P)\psi (x,P)$
and $\delta{\cal L} = \delta u(x)$ gives
\be\label{kpcase}
\varpi  =  - \int _{dx} \delta u(x) \sum _i
\left( {dp\over\langle \psi ^{\dagger}\psi\rangle}\psi ^{\dagger}(x)\psi (x)
\right) (\gamma _i)
\ee
The differential
$d\Omega = {dp\over\langle \psi ^{\dagger}\psi\rangle}
\psi ^{\dagger}(x)\psi (x)$ is
holomorphic on $\Sigma $ except for the ''infinity" point $P_0$ where it has
zero residue \cite{KriUMN}. Vanishing of the sum of residues of its
"special" variation
\footnote{It should be pointed out that the variation $\tilde\delta $
corresponds to a rather specific situation when one shifts only $\psi $
keeping $\psi ^{\dagger}$ fixed.}
\be\label{reskp}
{\tilde\delta}\left(\res _{P_0}d\Omega + \sum _i\res _{\gamma _i}d\Omega \right) = 0
\ee
can be rewritten as
\be\label{reskp2}
\delta v(x) + \sum _i d\Omega (\gamma _i) = 0
\ee
where $v(x)$ is a ''residue" of the BA function at the point $P_0$ obeying
$v'(x) = u(x)$. Substituting (\ref{reskp2}) into (\ref{general}) one gets
\be\label{kdv1}
\varpi = \int _{dx} \delta u(x) \int ^x _{dx'}\delta u(x')
\ee
or the {\it first} symplectic structure of the KdV equation.\\
{\bf Toda chain/lattice}. (The case directly related to the \N2 {\it
pure} SYM theory). One has
$$
\langle \psi ^{\dagger}\psi\rangle = \sum _n\psi ^+_n(P)\psi ^-_n(P)
$$
and the Lax equation acquires the form (\ref{laxtoda})
where $t = t_+ + t_-$ and $t_1 = t_+ - t_-$ is the first time of the Toda
chain, so that
\be\label{varEtoda}
\delta\lambda = {\sum _k \psi ^+_k \delta
p_k\psi ^-_k\over \langle \psi ^+\psi ^-\rangle}
\ee
and (\ref{general})
becomes
\be\label{toda}
\varpi  =  - \sum _k \delta p_k \sum _i \left(
{dp\over\langle \psi ^+\psi ^-\rangle}\psi ^+_k \psi ^-_k
\right) (\gamma _i)
\ee
Thus, to get
\be\label{ochain}
\varpi = \sum _k \delta p_k\wedge\delta q_k
\ee
one has to prove
\be\label{varx}
\sum _i \left(
{dp\over\langle \psi ^+\psi ^-\rangle}\psi ^+_k \psi ^-_k
\right) (\gamma _i)  = \delta q_k
\ee
To do this one considers again
\be\label{restoda}
{\tilde\delta} \left(\res _{P_+} + \res _{P_-} +
\sum _i\res _{\gamma _i}\right)d\Omega _n = 0
\ \ \ \ \ \
d\Omega _n =
{dp\over\langle \psi ^+\psi ^-\rangle}\psi ^+_n \psi ^-_n
\ee
where the first two terms for $\psi ^{\pm}_n \stackreb{\lambda
\rightarrow\lambda (P_{\pm})}{\sim} e^{\pm q_n}\lambda ^{\pm n}(1 +
{\cal O}(\lambda ^{-1}))$ satisfying two ''shifted" equations (\ref{laxtoda})
(with
${\tilde q}_n$ and $q_n $ correspondingly) give $\delta q_n = {\tilde q}_n -
q_n$ while the rest -- the l.h.s. of (\ref{varx}).

{\bf Calogero-Moser system}.
Introducing the ''standard" $dE$ and $dp$ one the curve $\Sigma $
(\ref{fscCal})
with the 1-form (\ref{dSCal})
where $dp=d\xi$ is holomorphic on torus $\oint _A dp = \omega$,
$\oint _B dp = \omega '$ and $E=\lambda $
has $N_c-1$ poles with $residue = 1$ and $1$ pole with $residue = -(N_c-1)$,
the BA function is defined by \cite{KriCal}
\be\label{bacal}
{\cal L}^{Cal}(\xi ){\bf a} =  \lambda {\bf a}
\ee
with the essential singularities
\be\label{baprop}
a_i \stackreb{E = E_+}{\sim} e^{x_i\zeta (\xi )}\left( 1 + {\cal O}(\xi)\right)
\ \ \ \ \
a_i \stackreb{E \neq E_+}{\sim} e^{x_i\zeta (\xi)}\left( -{1\over n-1} +
{\cal O}(\xi)\right)
\ee
and (independent of dynamical variables) poles ${\bf \gamma}$.  Hence,
similarly to the above case for the eq. (\ref{general}) one has
$\langle \psi ^{\dagger}\psi\rangle = \sum _ia_i^{\dagger}(P) a_i(P)$,
$\delta {\cal L}^{Cal} = {\sum _ia_i^{\dagger}(P) \delta p_i a_i(P)\over
\sum _ia_i^{\dagger}(P) a_i(P)}$ so that the expression (\ref{general})
acquires the form
\be\label{calo}
\varpi =  - \sum _k \delta p_k \sum _i \left(
{d\xi\over\langle a^{\dagger}a\rangle}a^{\dagger}_k a_k
\right) (\gamma _i)
\ee
and the residue formula
\be\label{rescal}
{\tilde\delta} \left(\sum _{P_j:p = 0}\res _{P_j} +
\sum _i\res _{\gamma _i}\right)d\Omega _k = 0
\ \ \ \ \ \ \
d\Omega _k =
{d\xi\over\langle a^{\dagger}a\rangle}a^{\dagger}_k a_k
\ee
where the first sum is over all ''infinities" $p = 0$ at each sheet of the
cover (\ref{fscCal}). After variation and using (\ref{baprop}) it gives
again
\be\label{calogero}
\varpi = \sum _k \delta p_k\wedge \delta x_k
\ee
The general proof of the more cumbersome analog of the above derivation
can be found in \cite{KriPho}. The explicit form of the obtained
formulas looks rather nontrivial, hence, to illustrate their existence
let us, finally, demonstrate the existence of (\ref{reskp}),
(\ref{restoda}) and (\ref{rescal}) for the 1-gap solution. Let
\be\label{psya}
\psi = e^{x\zeta (z)}
{\sigma (x-z+\kappa)\over\sigma (x+\kappa)\sigma (z-\kappa)}
\ \ \ \  \ \ \ \
\psi ^{\dagger}= e^{-x\zeta (z)}
{\sigma (x+z+\kappa)\over\sigma (x+\kappa)\sigma (z+\kappa)}
\ee
be solutions to
\be\label{1gap}
(\partial ^2 + u)\psi = (\partial ^2 - 2\wp (x+\kappa))\psi = \wp (z)\psi
\ee
Then
\be
\psi ^{\dagger}\psi = {\sigma (x-z+\kappa )\sigma (x+z+\kappa )\over\sigma ^2(x+\kappa )
\sigma (z-\kappa )\sigma (z+\kappa )}
= {\sigma ^2(z)\over\sigma (z +\kappa )\sigma (z - \kappa )}
\left( \wp (z) - \wp (x+\kappa )\right)
\nn \\
\langle \psi ^{\dagger}\psi\rangle =
= {\sigma ^2(z)\over\sigma (z +\kappa )\sigma (z - \kappa )}
\left( \wp (z) - \langle\wp (x+\kappa )\rangle \right)
\ee
and let us take the average over a period $2{\tilde\omega}$ to be
$\langle\wp (x+\kappa )\rangle = 2{\tilde\eta}$. Also
\be\label{qm}
dp = d\left(\zeta (z) + \log\sigma (2{\tilde\omega}-z+\kappa ) - \log\sigma
(\kappa - z)\right) = -dz\left(\wp (z) + \zeta (2{\tilde\omega}-z+\kappa ) -
\zeta (\kappa -z)\right) =
\nn \\
= -dz\left( \wp (z) - 2{\tilde\eta}\right)
\ee
and
\be\label{ppsi}
{dp\over\langle\psi ^{\dagger}\psi\rangle} =
dz{\sigma (z+\kappa )\sigma (z-\kappa)\over\sigma ^2(z)}
\nn \\
d\Omega = {dp\over\langle\psi ^{\dagger}\psi\rangle}\psi ^{\dagger}\psi =
dz{\sigma (x+z+\kappa )\sigma (x+z-\kappa)\over\sigma ^2(z)\sigma ^2(x+\kappa )} =
dz\left(\wp (z) - \wp (x+\kappa )\right)
\ee
Now, the variation $\tilde\delta$ explicitly looks as
\be\label{var1gap}
\tilde\delta d\Omega \equiv {dp\over\langle\psi ^{\dagger}\psi\rangle}
\psi ^{\dagger}_{\kappa}\psi _{\kappa + \delta\kappa} =
dz
{\sigma (z+\kappa )\sigma (z-\kappa)\sigma (x+z+\kappa)
\sigma (x-z+\kappa +\delta\kappa )
\over
\sigma ^2(z)\sigma (x+\kappa)\sigma (z+\kappa)
\sigma (x+\kappa +\delta\kappa )\sigma (z-\kappa -\delta\kappa )} =
\nn \\
= dz{\sigma (x+\kappa +z)\sigma (x+\kappa -z)\over\sigma ^2(z)\sigma ^2(x +\kappa )}
\left[ 1 + \delta\kappa\left(\zeta (x-z+\kappa ) + \zeta (z-\kappa ) -
\zeta (x +\kappa )\right) + {\cal O}\left( (\delta\kappa)^2\right)\right]
\nn \\
= dz\left(\wp (z) - \wp (x +\kappa )\right)
\left[ 1 + \delta\kappa\left(\zeta (x-z+\kappa ) + \zeta (z-\kappa ) -
\zeta (x +\kappa )\right) + {\cal O}\left( (\delta\kappa)^2\right)\right]
\ee
It is easy to see that (\ref{var1gap}) has non-zero residues at $z=0$ and
$z = \kappa $ (the residue at $z = x+\kappa $ is suppressed by
$\wp (z) - \wp (x +\kappa )$. They give
\be\label{res0}
\res _{z=0} \delta d\Omega \sim \delta\kappa \oint _{z\hookrightarrow 0}
dz\wp (z)
\left(\zeta (x-z+\kappa )+\zeta (z-\kappa )\right) \sim
\delta\kappa \oint _{z\hookrightarrow 0}\zeta (z) d\left(\zeta (x-z+\kappa )
+ \right.
\nn \\
\left. +\zeta (z-\kappa )\right)\sim
\delta\kappa\left(\wp (x +\kappa ) +\wp (\kappa )\right)\sim
\delta\left(\zeta (x +\kappa ) + \zeta (\kappa)\right)\equiv \delta v(x)
\ee
and
\be
\res _{z=\kappa }\delta d\Omega = \delta\kappa \left(\wp (\kappa )-\wp (x+\kappa )\right)
= d\Omega (\kappa )
\ee
which follows from the comparison to (\ref{ppsi}). Thus, we have checked
the existence of (\ref{reskp}) and (\ref{reskp2}) in the simplest
possible example by explicit computation.

Finally, the quantization of the symplectic structure (\ref{wsform}),
at least in the cases when it is known, corresponds to the complete
formulation of the effective theory. The corresponding generating function
$S = \sum _k\int ^{\gamma _k}dS$ defines the duality transformation
between the dual integrable systems (see \cite{KM1,KM2}). The formulation
of the complete effective theory hypothetically corresponds to the exact
answer for the full generating function $\log{\cal T} = \log{\cal T}_0
+ \log{\cal T}_{\theta} \equiv {\cal F}+ \log{\cal T}_{\theta}$,
including also the deformation of the oscillating part corresponding to
{\it massive} states in the spectrum.

\subsection{Comments on main definitions \label{invar}}

Let us now make some comments on the main formulas
(\ref{hol}) etc we used above. In fact these formulas need to be
defined in a more strict way since they imply taking derivatives
of various objects (holomorphic and meromorphic differentials on
complex curves) over moduli and {\it a priori} this is not a very
well-defined procedure. In most of concrete cases the hyperelliptic
curves were used (i.e. the curves which possess a special coordinate
$\lambda $ or differential $d\lambda $ with zero periods along any cycles),
and by derivatives over ("hyperelliptic") moduli -- or over the ramification
points one meant
\be\label{defder}
{\d\over\d h_k}dS \equiv \left.{\d\over\d h_k}dS(\lambda ,h_k)
\right|_{\lambda = const}
\ee
i.e. ${\d\over\d h_k}d\lambda = 0$. Now, one should consider at least
two arising questions:
\begin{itemize}
\item Why this definition is reasonable for hyperelliptic Riemann surfaces
(e.g. (\ref{fsc-Toda}), (\ref{caln2}), (\ref{fsc-sc2}))?
\item What should be done instead, when there is no specific coordinate or
differential like $d\lambda $?
\end{itemize}
In this subsection I would try at least partially to answer to this question.
These comments arose as a result of discussions with A.Levin and
his many clear explanations.

Consider the definition of an integrable system in spirit of \cite{DKN}
\be\label{hol1}
{\d\over\d h_k}dS \cong dv _k
\ee
where $dS$ is generating differential and $dv_k$ are {\em some} holomorphic
differentials. We will try to demonstrate how the generating differential
(\ref{hol}) can be constructed for a generic integrable system.

Consider, first, the purely holomorphic case (appearing for example
in the framework of the Hitchin systems \cite{Hi}). The most general
construction of the differentials of the (\ref{hol}) type is based on the
existence of a differentials with $g-1$ {\em double} zeroes on a curve
$\Sigma _g$ of genus $g$. These differentials look like
\be\label{dShol}
dS(P) \sim \sum _{j=1}^g{\d\Theta _{\ast}\left({\bf 0}\right)\over\d A_j}
d\omega _j (P) \equiv H_{\ast}(P) \equiv \nu _{\ast}^2(P)
\ee
where ${\bf A}(P) = \int ^Pd{\bomega}$ is the Abel map
into Jacobian ${\rm Jac}(\Sigma _g)$, $ \Theta$ --
Riemann theta-function \cite{Fay,Mumford},
$\nu _{\ast}(P)$ is a section of $K^{1\over 2}$ corresponding to an
even characteristic $\ast $ having simple zeroes at points
$R_1,\dots,R_{g-1}$. On a genus $g$ curve one has $2^{2g-1}+2^{g-1}$ even
and $2^{2g-1}-2^{g-1}$ odd characteristics.

Now one should define the co-ordinates $\{ h_k\}$ -- {\em locally} the
directions in moduli space such that the derivatives in this direction
give holomorphic differentials. The correspondence can be established as
\be\label{virk}
1 + \epsilon {\d\over\d h_k} \leftrightarrow 1 + \epsilon L^{(k)}_{-2}
\nn \\
L^{(k)}_{-2} = {1\over\xi_k}{\d\over\d\xi_k}
\ee
where $\xi_k$ is a co-ordinate in the vicinity of a point $R_k$. They
obviously commute
\be\label{comm}
\left[ L^{(j)}_{-2},L^{(k)}_{-2}\right] = 0
\ee
An easy way
to check the correspondence is to consider the following basis in the space of
holomorphic 1-differentials on $\Sigma _g$ \cite{BeMa}:
\be
dv_k \stackreb{P\rightarrow R_k}{\sim} d\xi _k
\stackreb{P\rightarrow R_j}{+} \delta _{jk}\xi _j d\xi _j
\nn \\
dv_g = dS \stackreb{P\rightarrow R_j}{\sim}\xi _j^2d\xi
\nn \\
j\neq k \in \{ 1,\dots,g-1 \}
\ee
It is possible to write down explicit formulas for the differentials $dv_k$.
To do this one should first write down a section of $K^{1\over 2}$ with the
only simple pole at $R_j$. It has the following form
\be\label{fermpole}
\psi _j(P) \sim {{\d\over\d A_i}\Theta _{\ast}({\bf A}(P)-{\bf A}(R_j))\over
{\d\over\d A_i}\Theta _{\ast}({\bf 0})\ E(P,R_j)}
\nn \\
\psi _j(P) \stackreb{P\rightarrow R_j}{\sim} {\sqrt{d\xi _j}\over\xi _j}
+ \dots
\ee
where we used explicitely the Prime form
\be\label{szego}
E(P,P_0) = {\Theta _{\ast}({\bf A}(P) - {\bf A}(P_0))\over \nu (P)\nu (P_0)}
\ee
Then, obviously
\be
dv_k(P) = \psi _k (P)\nu _{\ast}(P)
\ee
Now, to check (\ref{hol}) one acts by generators (\ref{virk}) to the
$dS = dv_g$. The result obviously is
\be
\left( 1 + \epsilon {\d\over\d h_i}\right) dS =
\left( 1 + \epsilon L^{(i)}_{-2}\right)dS =
\nn \\
= \left( 1 + \epsilon {1\over\xi_i}{\d\over\d\xi_i}\right)
(\xi _i^2d\xi _i + \dots) = \left(\xi _i + {\epsilon\over\xi _i}\right)^2
d\left(\xi _i + {\epsilon\over\xi _i}\right) =
\nn \\
= \xi _i^2d\xi _i + \epsilon d\xi _i + {\cal O}(\epsilon ^2) + \dots =
dS + \epsilon dv_k + {\cal O}(\epsilon ^2)
\ee
i.e. indeed the equality (\ref{hol}). To get $2g$-dimensional integrable
system one could add as a parameter the normalization of $dS$, i.e.
\be
dS = h_gdv_g
\ee
The above considerations can be almost literally transformed to the
more interesting meromorphic case. The corresponding formulas can be
extracted from \cite{Fay}.

\subsection{Whitham equations}

Now, let us turn to another
definition of the generating 1-form (\ref{dS}). This definition goes back
to the general approach to
construction of the effective actions which is known as the
Bogolyubov-\-Whitham
averaging method (see \cite{TS,KriW,Kri,DN} for a comprehensive review
and references).  Though the Whitham dynamics is describes the commutative
flows on the moduli spaces, averaging over the Jacobian -- the fast part
of the theory,
its explicit formulation is most simple and natural in terms of
connections on spectral curves \cite{KriW,Kri}. The convenience of
the Whitham language is caused, in part, by the fact that the previous
consideration does not lead to any natural definition of the prepotential
${\cal F}$. The Whitham dynamics allows one to define generating differential
(\ref{hol}) in such a way that a natural
identification of (\ref{prepot}) with the {\em logariphm of
$\tau $-\-function} of the Whitham hierarchy appears.

The Whitham equations determine a flat coordinate system on some
(finite-dimensional, complex) space \cite{KriW}, which usually
appears in interesting examples as the space of complex structures
of a Riemann surface, associated to a finite-gap solution to the
equations of KP/Toda type.
The most part of interesting solutions to the Whitham hierarchy is related
to the ''modulation"
of parameters of the finite-gap solutions of integrable systems of KP/Toda type.
The KP $\tau$-function
associated with a given spectral curve $\Sigma _g$ is
\be\label{KPsol}
{\cal T}\{t_i\} = e^{\sum t_i\gamma _{ij}t_j}\Theta\left({\bf \Phi}_0
+
\sum t_i{\bf k}_i\right) \ \ \
{\bf k}_i = \oint_{\bf B} d\Omega_i
\ee
where $ \Theta$ is a Riemann theta-function \cite{Fay,Mumford} -- a function
on the Jacobian ${\rm Jac}(\Sigma _g)$ and $d\Omega_i$ are
meromorphic 1-differentials with poles
of the order $i+1$ at a marked point $z_0$. They are completely
specified by the normalization relations
\be
\oint_{\bf A} d\Omega_i = 0
\label{norA}
\ee
and the asymptotic behavior
\be
d\Omega_i = \left(\xi^{-i-1} + o(\xi)\right) d\xi
\label{norc}
\ee
where $\xi$ is a local coordinate in the vicinity of $z_0$.
The moduli $\{ u_\alpha \}$ of spectral curve are
invariants of KP flows,
\be
\frac{\partial u_\alpha}{\partial t_i} = 0,
\ee
However, the moduli become
dependent on $t_i$ after the ''modulation" defined  by the
Whitham equations which for a particular choice of the coordinate
on $\Sigma $ acquire the most simple form
\be\label{zwhi}
\frac{\partial d\Omega_i(z)}{\partial t_j} =
\frac{\partial d\Omega_j(z)}{\partial t_i}.
\ee
and imply that
\be\label{SS}
d\Omega_i(z) = \frac{\partial dS(z)}{\partial t_i}
\ee
and the equations for moduli, following from (\ref{zwhi}), are:
\be\label{whiv}
\frac{\partial u_\alpha}{\partial t_i} =
v_{ij}^{\alpha\beta}(u)\frac{\partial u_\beta}{\partial t_j}
\ee
with some (in general complicated) functions $v_{ij}^{\alpha\beta}$.

From the point of view of subsect.\ref{invar} the Whitham equations
relate two different set of (flat) coordinates on moduli space. One is
generated
by "averaged" KP/Toda flows and is given by $L_{-n}^{0}$ -- the singular
Virasoro generators in the pucture $P_0$ corresponding to the KP
hierarchy from geometrical point of view. The other are coordinates induced
by $L_{-2}^{\alpha}$ -- in the particular points $R_{\alpha}$ on the curve
$\Sigma $ which are branch points (or Riemann invariants) in the hyperelliptic
case.

In the KdV (and Toda-chain) case  all the spectral curves are
hyperelliptic, and for the KdV $i$ takes only odd
values $i = 2j+1$, so that
\be
d\Omega_{2j+1}(z) = \frac{{\cal P}_{j+g}(z)}{y(z)}dz,
\ee
the coefficients of the polynomials ${\cal P}_j$ being fixed
by normalization conditions (\ref{norA}), (\ref{norc})
(one usually takes $z_0 = \infty$ and the local parameter
in the vicinity of this point is $\xi = z^{-1/2}$). In this case the
equations (\ref{whiv}) can be diagonalized if the co-ordinates
$ \left\{  u_\alpha \right\}$ on the moduli space are taken to be the
ramification points:
\begin{equation}\label{whihy}
v_{ij}^{\alpha\beta}(u) = \delta ^{\alpha\beta}\left.{d\Omega
_i(z)\over d\Omega _j(z)}\right| _{z=u_\alpha}
\end{equation}
Finally, the differential $dS(z)$ (\ref{SS}) can be constructed for any
finite-\-gap solution \cite{DKN} and it {\it coincides} with the generating
1-form (\ref{dS}). The equality
\be\label{defF}
\frac{\partial {\cal F}}{\partial {\bf a}} = {\bf a}_D \ \ \ \ \ \ \
{\bf a} = \oint_{\bf A} dS \ \ \ \ \ \ \ \ \ {\bf a}_D = \oint_{\bf B} dS
\ee
defines $\tau $-\-function of the Whitham hierarchy
${\cal F} = \log {\cal T}_{Whitham}$ \cite{KriW,Kri}.
The exact answer for the partition function $\log{\cal T} = \log{\cal T}_0
+ \log{\cal T}_{\theta} \equiv {\cal F}+ \log{\cal T}_{\theta}$
should also include the deformation of the oscillating
part, corresponding to the {\it massive} excitations.
Below, the explicit examples
of the Whitham solutions will be considered.

Now let us demonstrate that the higher genus Riemann surfaces
(already in the elliptic case) give
rise to nonperturbative formulation of physically less trivial theories.
In contrast to the previous example Whitham times will be nontrivially
related to the moduli of the curve.
The elliptic solution to the KdV is
\be\label{ukdvsol}
U(t_1,t_3,\ldots| u) =
\frac{\partial^2}{\partial t_1^2} \log{\cal T} (t_1,t_3,\ldots|u) =
\\
= U_0 \wp (k_1t_1 + k_3t_3 + \ldots + \Phi_0 |\omega , \omega ') +
 {u\over 3}
\ee
and
\be
dp \equiv d\Omega_1(z) = \frac{z - \alpha (u)}{y(z)}dz, \nn \\
dQ \equiv d\Omega_3(z) = \frac{z^2 - \frac{1}{2}uz - \beta
(u)}{y(z)}dz.
\label{pp}
\ee
Normalization conditions (\ref{norA}) prescribe that
\be
\alpha (u) = \frac{\oint_A \frac{zdz}{y(z)}}
{\oint_A \frac{dz}{y(z)}}\ \ \ \ \
\beta (u) = \frac{\oint_A \frac{(z^2 - \frac{1}{2}uz)dz}{y(z)}}
{\oint_A \frac{dz}{y(z)}}.
\ee
The simplest elliptic example is the first Gurevich-Pitaevsky (GP) solution
\cite{GP} with the underlying spectral curve
\beq\label{elc1}
y^2 = (z^2 - 1)(z-u)
\eeq
specified by a requirement that all branching points except for $z = u$ are
{\it fixed} and do not obey Whitham deformation.
It is easy to see that by change of variables $z = u + \lambda ^2$
and $y \rightarrow y\lambda$ the curve (\ref{elc1}) can be written as
(a particular $N_c = 2$ case) of the Toda-\-chain curve (\ref{hypelTC})
$y^2 = (\lambda^2 + u)^2 - 1$ c $P_{N_c=2}(\lambda ) = \lambda^2 + h_2$ i.e.
$h_2 \equiv u$.
The generating differential (\ref{dS}), (\ref{SS}),
corresponding to (\ref{elc1}) is given by \cite{GKMMM}
\be\label{dSz}
dS(z) = \left(t_1 + t_3(z+\frac{1}{2}u)  + \ldots \right)
\frac{z-u}{y(z)}dz \ \stackreb{\{ t_{k>1}=0 \} }{=} \
t_1 \frac{z-u}{y(z)}dz
\ee
and it produces the simplest solution to (\ref{zwhi}) coming from the elliptic
curve. From (\ref{dSz}) one derives:
\be
\frac{\partial dS(z)}{\partial t_1} =
\left( z - u - (\frac{1}{2}t_1 + u t_3)
\frac{\partial u}{\partial t_1}\right)\frac{dz}{y(z)}, \nn \\
\frac{\partial dS(z)}{\partial t_3} =
\left( z^2 - \frac{1}{2}uz - \frac{1}{2}u^2 -
(\frac{1}{2}t_1 + u t_3)
\frac{\partial u}{\partial t_3}\right)\frac{dz}{y(z)}, \nn \\
\ldots ,
\ee
and comparison with explicit expressions (\ref{pp}) implies:
\be
(\frac{1}{2}t_1 + u t_3)\frac{\partial u}{\partial t_1} =
\alpha (u) - u, \nn \\
(\frac{1}{2}t_1 + u t_3)\frac{\partial u}{\partial t_3} =
\beta (u) - \frac{1}{2}u^2.
\label{preW}
\ee
In other words, this construction provides the first GP solution to the
Whitham equation
\be\label{GPe}
\frac{\partial u}{\partial t_3} = v_{31}(u)
\frac{\partial u}{\partial t_1},
\ee
with
\be
v_{31}(u) = \frac{\beta(u)-\frac{1}{2}u^2}{\alpha(u)-u} =
\left.\frac{d\Omega_3(z)}{d\Omega_1(z)}\right|_{z=u},
\ee
which can be expressed through elliptic integrals \cite{GP}.
More detailed analysis can be found in \cite{M4}.

\section{Prepotential of the Effective Theory}

In the previous section the prepotential ${\cal F}$ was identified with the
logariphm of the $\tau $-\-function of the Whitham hierarchy. Such
identification, being a particular case of the fundamental formula relating
the generating functions or effective actions of quantum theories to the
$\tau $-\-functions of the hierarchies of integrable equations, has a little
bit implicit form. In this section, I discuss an
explicit system of differential equations satisfied by the prepotential
${\cal F}$. These are the associativity equations \cite{MMM2,MMM3,MMM4}
and their existence for the prepotential follows, in principle, from the
fact that it is $\tau$-function of the Whitham hierarchy (though the
associativity equations exist even in more general situation
\cite{Dubtop,Manin}). More exactly, in this section I will formulate the most
general form of the associativity equations and check their existence
in $2d$ topological models and in the effective Seiberg-Witten theories.
Thus, we will deal with the explicit differential equations having
among their dolutions the prepotentials ${\cal F}$.

\subsection{The associativity equations}

The prepotential ${\cal F}$ \cite{SW1,SW2} is defined in terms of a
family of Riemann surfaces,
endowed with the meromorphic differential
$dS$.  For the gauge group $G=SU(N)$ the family is
\cite{SW1,SW2,sun,GKMMM} given by (\ref{fsc-Toda})
and the generating differential by (\ref{dS}).
The prepotential ${\cal F}(a_i)$ is implicitly defined by the set of
equations (\ref{defF}).
According to \cite{GKMMM}, this definition identifies
${\cal F}(a_i)$ as logarithm of (truncated) $\tau$-function of
Whitham integrable hierarchy.
Existing experience with Whitham hierarchies
\cite{KriW,Dubtop}
implies that ${\cal F}(a_i)$ should satisfy some sort
of the Witten-\-Dijkgraaf-\-Verlinde-\-Verlinde (WDVV) equations
\cite{WDVV}.
Below in this section we demonstrate that WDVV equations
for the prepotential actually look like \cite{MMM2}
\be
{\cal F}_i {\cal F}_k^{-1} {\cal F}_j = {\cal F}_j {\cal F}_k^{-1} {\cal F}_i
\ \ \ \ \ \ \forall i,j,k = 1,\ldots,N-1.
\label{FFF}
\ee
Here ${\cal F}_i$ denotes the matrix
\be
({\cal F}_i)_{mn} = \frac{\partial^3 {\cal F}}{\partial a_i
\partial a_m\partial a_n}.
\ee
Let us first present few comments about equations (\ref{FFF}):
\begin{itemize}
\item Let us, first, notice that the conventional WDVV equations
for topological ($2d$) field theory express the associativity of the algebra
$\phi_i\phi_j = C_{ij}^k\phi_k$ (for symmetric in $i$ and $j$ structure
constants):
$(\phi_i\phi_j)\phi_k = \phi_i(\phi_j\phi_k)$,
or $C_iC_j = C_jC_i$, for the matrices  $(C_i)^m_n \equiv C_{in}^m$.
These conditions become highly non-trivial since, in
topological theory, the structure constants are expressed
in terms of a single prepotential ${\cal F}(t_i)$:
$C_{ij}^l = (\eta^{-1}_{(0)})^{kl}{\cal F}_{ijk}$, and
${\cal F}_{ijk} = {\partial^3{\cal F}\over\partial t_i\partial t_j\partial t_k}$,
while the metric is $\eta_{kl}^{(0)} = {\cal F}_{0kl}$, where $\phi_0 = I$
is the unity operator. In other words, the conventional WDVV equations
can be written as
\be
{\cal F}_i {\cal F}_0^{-1} {\cal F}_j = {\cal F}_j {\cal F}_0^{-1} {\cal F}_i.
\label{FFFconv}
\ee
In contrast to (\ref{FFF}), in the standard WDVV equations
to $k=0$, associated with the distinguished unity operator.

On the other hand, in the Seiberg-Witten effective theory there is no
distinguished index $i$: all the arguments $a_i$ of the
prepotential can be treated on equal footing. Thus, if some kind of
the WDVV equations holds in this case, it should be invariant under
any permutation of indices $i,j,k$ -- criterium satisfied by
the system (\ref{FFF}).
Moreover, the same set of equations (\ref{FFF}) is satisfied for
generic topological theory.
\item In the general theory of Whitham hierarchies
\cite{KriW,Dubtop} the WDVV equations arise also in the form (\ref{FFFconv}).
Again, there exists  a distinguished time-variable
for the global solutions usually associated with the first time-variable
of the original
KP/KdV hierarchy. Moreover, usually -- in contrast to the
simplest topological models -- the set of these variables for
the Whitham hierarchy is infinitely large. In this context
our eqs.(\ref{FFF}) state that, for specific subhierarchies
(in the Seiberg-Witten gluodynamics, it is the Toda-chain hierarchy,
associated with a peculiar set of hyperelliptic surfaces),
there exists a non-trivial {\it truncation} of the
quasiclassical $\tau$-function, when it depends on the finite
number ($N-1=g$ -- genus of the Riemann surface) of
{\it equivalent} arguments $a_i$, and satisfies a much wider
set of WDVV-like equations: the whole set (\ref{FFF}).
\item From (\ref{defF}) it is clear that $a_i$'s
are defined modulo linear transformations
(one can change $A$-cycle for any linear combination
of them). Eqs.(\ref{FFF}) possess adequate ``covariance'':
the least trivial part is that ${\cal F}_k$ can be
substituted by ${\cal F}_k + \sum_l\epsilon_l {\cal F}_l$. Then
\be\label{F-1}
{\cal F}_k^{-1} \rightarrow ({\cal F}_k + \sum \epsilon_l{\cal F}_l)^{-1}
= {\cal F}_k^{-1} - \sum\epsilon_l {\cal F}_k^{-1}{\cal F}_l{\cal F}_k^{-1} +
\sum \epsilon_l\epsilon_{l'} {\cal F}_k^{-1}{\cal F}_l{\cal F}_k^{-1}
{\cal F}_{l'}
{\cal F}_k^{-1} + \ldots
\ee
Clearly, (\ref{FFF}) -- valid for all
triples of indices {\it simultaneously} -- is enough to guarantee that
${\cal F}_i({\cal F}_k + \sum \epsilon_l{\cal F}_l)^{-1}{\cal F}_j =
{\cal F}_j({\cal F}_k + \sum \epsilon_l{\cal F}_l)^{-1}{\cal F}_i$.
Covariance under any replacement of $A$ and $B$-cycles together will be
seen from the general proof below: in fact the role
of ${\cal F}_k$ can be played by ${\cal F}_{d\omega}$, associated with
{\it any} holomorphic 1-differential $d\omega$ on the
Riemann surface.
\item The metric $\eta$, which is a second derivative,
(as is the case for our $\eta^{(k)}_{mn}\equiv ({\cal F}_k)_{mn}$)
in ordinary topological theories ($\eta^{(0)}$) is always flat,
and this allows one to choose ``flat coordinates'' where
$\eta^{(0)} = const$. Sometimes {\it all} the metrics $\eta^{(k)}$
are flat simultaneously. However, this is not always the case: in the
example of quantum cohomologies of $CP^2$ \cite{KoMa,Manin} eqs.(\ref{FFF})
are true for all $k=0,1,2$, but only $\eta^{(0)}$
is flat while $\eta^{(1)}$ and
$\eta^{(2)}$ lead to non-vanishing curvatures.

The equations (\ref{FFF}) are {\it trivially} satisfied in the case of
$N=2$ and $N=3$ and become a nontrivial condition only starting with $N\geq 4$.

\item Our consideration suggests that when
the {\it ordinary} WDVV (\ref{FFFconv}) is true, the whole system (\ref{FFF})
holds automatically for any other $k$ (with the only restriction that
${\cal F}_k$ is non-\-degenerate).
Indeed,
\footnote{This simple proof was suggested by A.Rosly.}
\be
{\cal F}_i{\cal F}_k^{-1}{\cal F}_j =
{\cal F}_0 \left(C_i^{(0)}(C_k^{(0)})^{-1} C_j^{(0)}\right)
\ee
is obviously symmetric w.r.to the permutation $i\leftrightarrow j$ implied
by $[C_i^{(0)},C_j^{(0)}] = 0$.
\item Effective theory (\ref{defF}) is naively {\it non-\-topological}.
From the 4-dimensional point of view it describes the
low-energy limit of the Yang-Mills theory which -- at least, in
the ${\cal N}=2$ supersymmetric case -- is {\it not} topological and
contains propagating massless particles. Still this theory
is entirely defined by a prepotential, which -- as we now
see -- possesses {\it all} essential properties of the
prepotentials in topological theory. Thus, from the
``stringy'' point of view (when everything is described
in terms of universality classes of effective actions)
the Seiberg-Witten models belong to the same class as
topological models: only the way to extract physically
meaningful correlators from the prepotential is
different. This can serve as a new evidence
that the notion of topological theories is deeper than
it is usually assumed: as emphasized in \cite{GKMMM} it
can be actually more related to the low-energy (IR) limit of
field theories than to the property of the correlation
functions to be constants in physical space-time.

Moreover, the fact that only {\em third} derivatives enter the
equation (\ref{FFF}) demonstrates the stringy origin
of the nonperturbative solutions {\it a la} Seiberg-Witten.
\end{itemize}

\subsection{The proof of the associativity equations}

Let us start with reminding the proof of the WDVV equations
(\ref{FFFconv}) for ordinary topological theories.
We take the simplest of all possible examples, when
$\phi_i$ are polynomials of a single variable $\lambda$.
The proof is essentially the check of consistency between the
following formulas:
\be
\phi_i(\lambda)\phi_j(\lambda) = C_{ij}^k\phi_k(\lambda)
\ {\rm mod}\ W'(\lambda),
\label{.c}
\ee
\be
{\cal F}_{ijk} = {\rm res}\frac{\phi_i\phi_j\phi_k(\lambda)}
{W'(\lambda)} \equiv \sum_{\alpha}
\frac{\phi_i\phi_j\phi_k(\lambda_\alpha)}
{W''(\lambda_\alpha)},
\label{vc}
\ee
\be
\eta_{kl} = {\rm res}\frac{\phi_k\phi_l(\lambda)}
{W'(\lambda)} \between \sum_{\alpha}
\frac{\phi_k\phi_l(\lambda_\alpha)}
{W''(\lambda_\alpha)},
\label{vvc}
\ee
\be
{\cal F}_{ijk} = \eta_{kl}C_{ij}^l.
\label{vvvc}
\ee
Here $\lambda_\alpha$ are the roots of $W'(\lambda)$.

In addition to the consistency of (\ref{.c})-(\ref{vvvc}),
one should know that {\it such} ${\cal F}_{ijk}$, given by
(\ref{vc}), are the third derivatives of a single function ${\cal F}(a)$, i.e.
\be
{\cal F}_{ijk} = \frac{\partial^3{\cal F}}{\partial a_i
\partial a_j\partial a_k}.
\ee
This integrability property of (\ref{vc}) follows from separate
arguments and can be checked independently.
But if (\ref{.c})-(\ref{vvc}) is given, the proof of
(\ref{vvvc}) is straightforward:
\be \eta_{kl}C^l_{ij} = \sum_{\alpha}
\frac{\phi_k\phi_l(\lambda_\alpha)}
{W''(\lambda_\alpha)} C^l_{ij} \stackrel{(\ref{.c})}{=} \\ =
\sum_{\alpha}
\frac{\phi_k(\lambda_\alpha)}
{W''(\lambda_\alpha)} \phi_i(\lambda_\alpha)\phi_j(\lambda_\alpha)
= {\cal F}_{ijk}.
\ee
Note that (\ref{.c}) is defined modulo $W'(\lambda)$,
but $W'(\lambda_\alpha) = 0$ at all the points $\lambda_\alpha$.
Imagine now that we change the definition of the metric:
\be
\eta_{kl} \rightarrow \eta_{kl}(Q) =
\sum_{\alpha}
\frac{\phi_k\phi_l(\lambda_\alpha)}
{W''(\lambda_\alpha)}Q(\lambda_\alpha).
\ee
Then the WDVV equations would still be correct, provided the
definition (\ref{.c}) of the algebra is also changed for
\be
\phi_i(\lambda)\phi_j(\lambda) = C_{ij}^k(Q)\phi_k(\lambda)
Q(\lambda)\ {\rm mod}\ W'(\lambda).
\label{..c}
\ee
This describes an associative algebra, whenever the
polynomials $Q(\lambda)$ and $W'(\lambda)$ are co-prime,
i.e. do not have common divisors.
Note that (\ref{vc}) -- and thus the fact that ${\cal F}_{ijk}$
is the third derivative of the same ${\cal F}$ -- remains intact!
One can now take for $Q(\lambda)$ any of the operators
$\phi_k(\lambda)$, thus reproducing eqs.(\ref{FFF}) for
all topological theories
\footnote{To make (\ref{FFF})
sensible, one should require that $W'(\lambda)$ has only
{\it simple} zeroes, otherwise some of the matrices ${\cal F}_k$
can be degenerate and non-invertible.}.

In the case of the Seiberg-Witten model the polynomials
$\phi_i(\lambda)$ are substituted by the canonical holomorphic
differentials $d\omega_i(\lambda )$ on hyperelliptic surface
(\ref{fsc-Toda}). This surface
can be represented in a standard hyperelliptic form (\ref{hypelTC}).
Instead of (\ref{.c}) and (\ref{..c}) we now put
\be
d\omega_i(\lambda )d\omega_j(\lambda ) =
C_{ij}^k(d\omega ) d\omega_k(\lambda )
d\omega (\lambda ) \ {\rm mod}\ \frac{dP_N(\lambda )d\lambda }{y^2}.
\label{.}
\ee
In contrast to (\ref{..c}) we can not now choose $Q = 1$
(to reproduce (\ref{.c})), because now we need it to be
a 1-differential. Instead we just take $d\omega $ to be a
{\it holomorphic} 1-differential. However, there is no distinguished
one -- just a $g$-parametric family -- and $d\omega $ can be
{\it any} one from this family. We require only that it is
co-prime with $\frac{dP_N(\lambda )}{y}$.

If the algebra (\ref{.}) exists, the structure constants
$C_{ij}^k(d\omega )$ satisfy the associativity condition
(if $d\omega $ and
${dP_N\over y}$ are co-prime). But we still need to show that
it indeed exists, i.e. that if $d\omega $ is given, one can find
($\lambda $-independent) $C_{ij}^k$. This is a simple exercise:
all $d\omega_i$ are linear combinations of
\be
dv_k(\lambda ) = \frac{\lambda ^{k-1}d\lambda }{y}, \ \ \ k=1,\ldots,g: \\
dv_k(\lambda ) = \sigma_{ki}d\omega_i(\lambda ), \ \ \
d\omega_i = (\sigma^{-1})_{ik}dv_k, \ \ \
\sigma_{ki} = \oint_{A_i}dv_k,
\label{sigmadef}
\ee
also $d\omega (\lambda ) = s_kdv_k(\lambda )$.
Thus, (\ref{.}) is in fact a relation between the polynomials:
\be
\left(\sigma^{-1}_{ii'}\lambda ^{i'-1}\right)
\left( \sigma^{-1}_{jj'}\lambda ^{j'-1}\right) =
C_{ij}^k \left(\sigma^{-1}_{kk'}\lambda ^{k'-1}\right)
\left( s_l\lambda ^{l-1}\right) +
p_{ij}(\lambda )P'_N(\lambda ).
\ee
At the l.h.s. we have a polynomial of degree $2(g-1)$.
Since $P'_N(\lambda )$ is a polynomial of degree $N-1=g$, this
implies that $p_{ij}(\lambda )$ should be a polynomial of degree
$2(g-1)-g = g-2$. The identification of two polynomials of
degree $2(g-1)$ impose a set of $2g-1$ equations for the coefficients.
We have a freedom to adjust $C_{ij}^k$ and $p_{ij}(\lambda )$
(with $i,j$ fixed), i.e. $g + (g-1) = 2g-1$ free parameters:
exactly what is necessary. The linear system of equations
is non-degenerate for co-prime $d\omega $ and $dP_N/y$.

Thus, we proved that the algebra (\ref{.}) exists (for a given
$d\omega $) -- and thus $C_{ij}^k(d\omega )$ satisfy the
associativity condition
\be
C_i(d\omega )C_j(d\omega ) = C_j(d\omega ) C_i(d\omega ).
\ee
Hence, instead of (\ref{vc}) we have \cite{KriW,Dubtop,Manin}:
\be
{\cal F}_{ijk} = \frac{\partial^3{\cal F}}{\partial a_i\partial a_j
\partial a_k} = \frac{\partial T_{ij}}{\partial a_k} = \nn \\
= \stackreb{d\lambda =0}{{\rm res}} \frac{d\omega_id\omega_j
d\omega_k}{d\lambda\left(\frac{dw}{w}\right)} =
\stackreb{d\lambda =0}{{\rm res}} \frac{d\omega_id\omega_j
d\omega_k}{d\lambda\frac{dP_N}{y}} =
\sum_{\alpha} \frac{\hat\omega_i(\lambda_\alpha)\hat\omega_j
(\lambda_\alpha)\hat\omega_k(\lambda_\alpha)}{P'_N(\lambda_\alpha)
/\hat y(\lambda_\alpha)}
\label{v}
\ee
The sum at the r.h.s. goes over all the $2g+2$ ramification points
$\lambda_\alpha$ of the hyperelliptic curve (i.e. over the zeroes
of $y^2 = P_N^2(\lambda )-1 = \prod_{\alpha=1}^N(\lambda - \lambda_\alpha)$);
\  $d\omega_i(\lambda) = (\hat\omega_i(\lambda_\alpha) +
O(\lambda-\lambda_\alpha))d\lambda$,\ $\ \ \ \hat y^2(\lambda_\alpha) =
\prod_{\beta\neq\alpha}(\lambda_\alpha - \lambda_\beta)$.
Formula (\ref{v}) can be extracted from \cite{KriW}, and its proof
is presented, for example, in \cite{MMM2}.

We define the metric in the following way:
\be
\eta_{kl}(d\omega ) =
\stackreb{d\lambda =0}{{\rm res}} \frac{d\omega_kd\omega_l
d\omega }{d\lambda\left(\frac{dw}{w}\right)} =
\stackreb{d\lambda =0}{{\rm res}} \frac{d\omega_kd\omega_l
d\omega_k}{d\lambda\frac{dP_N}{y}} = \\ =
\sum_{\alpha} \frac{\hat\omega_k(\lambda_\alpha)\hat\omega_l
(\lambda_\alpha)\hat Q(\lambda_\alpha)}{P'_N(\lambda_\alpha)
/\hat y(\lambda_\alpha)}
\label{vv}
\ee
In particular, for $d\omega  = d\omega_k$,
$\eta_{ij}(d\omega_k) = {\cal F}_{ijk}$: this choice will
give rise to (\ref{FFF}).

Given (\ref{.}), (\ref{v}) and (\ref{vv}), one can check:
\be
{\cal F}_{ijk} = \eta_{kl}(d\omega )C_{ij}^k(d\omega ).
\label{vvv}
\ee
Note that ${\cal F}_{ijk} = {\partial^3{\cal F}\over\partial a_i\partial a_j
\partial a_k}$ at the l.h.s. of (\ref{vvv}) is independent
of $d\omega $! The r.h.s. of (\ref{vvv}) is equal to:
\be
\eta_{kl}(d\omega )C_{ij}^k(d\omega ) =
\stackreb{d\lambda =0}{{\rm res}} \frac{d\omega_kd\omega_l
d\omega }{d\lambda\left(\frac{dw}{w}\right)} C_{ij}^l(d\omega )
\stackrel{(\ref{.})}{=} \\ =
\stackreb{d\lambda =0}{{\rm res}} \frac{d\omega_k}
{d\lambda\left(\frac{dw}{w}\right)}
\left(d\omega_id\omega_j - p_{ij}\frac{dP_Nd\lambda}{y^2}\right) =
{\cal F}_{ijk} - \stackreb{d\lambda =0}{{\rm res}} \frac{d\omega_k}
{d\lambda\left(\frac{dP_N}{y}\right)}p_{ij}(\lambda)\frac{dP_Nd\lambda}{y^2}
= \\ = {\cal F}_{ijk} - \stackreb{d\lambda =0}{{\rm res}}
\frac{p_{ij}(\lambda )d\omega_k(\lambda)} {y}
\ee
It remains to prove
that the last item is indeed vanishing for any $i,j,k$.
This follows from the
fact that $\frac{p_{ij}(\lambda )d\omega_k(\lambda )}{y}$
is singular only at zeroes of $y$, it is not singular at
$\lambda =\infty$ because
$p_{ij}(\lambda)$ is a polynomial of low enough degree
$g-2 < g+1$. Thus the sum of its residues at ramification points
is thus the sum over {\it all} the residues and therefore vanishes.

This completes the proof of associativity equations for the pure
\N2 SUSY Yang-Mills theory or the Toda chain integrable model \cite{MMM2}.
Taking $d\omega  = d\omega_k$ (which is obviously co-\-prime with
$\frac{dP_N}{y}$), we obtain (\ref{FFF}).

\subsection{Algebraic construction of WDVV equations}

Let us now discuss in detail the algebraic structure underlying the equations
(\ref{FFF}) \cite{MMM4,MMM3}. Remind, first, that for any metric
\be
G = \sum_m g^{(m)}{\cal F}_m
\label{metric}
\ee
used to raise up indices
\be
C^{(G)}_j = G^{-1}{\cal F}_j,
\label{defC}
\ee
i.e. $C^i_{jk} = (G^{-1})^{im}{\cal F}_{mjk}$, or
${\cal F}_{ijk} = G_{im}C^m_{jk}$
\footnote{From now on we omit the
superscript $(G)$ in $C^{(G)}$ and assume summation over
repeated indices.} the WDVV eqs imply that all matrices
$C$ commute:
\be
C_iC_j = C_jC_i \ \ \ \
\forall i,j
\label{comC}
\ee
(and thus can be diagonalized simultaneously).
While (\ref{FFF}) implies (\ref{comC}), inverse is not true:
the WDVV equations are either (\ref{FFF}) or the combination of
(\ref{metric}), (\ref{defC}) and (\ref{comC}).
Let us also remind that the WDVV eqs
(\ref{comC}) expresses the associativity of the multiplication of
observables $\phi_i$
(for example in the chiral rings \cite{chr} in $2d$ $N=2$ superconformal
topological models), where
\be
\phi_i \circ \phi_j = C^k_{ij} \phi_k, \\
(\phi_i\circ\phi_j)\circ\phi_k = \phi_i\circ(\phi_j\circ\phi_k),
\label{alg}
\ee
while
\be
{\cal F}_{ijk} = \langle\langle \phi_i \phi_j \phi_k \rangle\rangle
\ee
are (deformed) 3-point correlation functions on sphere.

The basic example of the algebra (\ref{alg}) is the multiplication
of polynomials modulo $dP$
\be\label{algp}
\phi_i(\lambda)\phi_j(\lambda) = C^k_{ij} \phi_k(\lambda)G'(\lambda)
\ {\rm mod}\ P'(\lambda)
\ee
Here $P(\lambda)$ and $G(\lambda)$
are polynomials of $\lambda$,
such that their $\lambda$-derivatives $P'(\lambda)$
and $G'(\lambda)$  are co-prime (do not have common divisors).
The algebra (\ref{algp}) is obviously associative as a factor of explicitly
associative multiplication algebra of polynomials over its ideal
$P'(\lambda) = 0$.

The second ingredient of the WDVV eqs is the residue formula \cite{ref},
\be
{\cal F}_{ijk} = \stackreb{dP = 0}{\res}\
\frac{\phi_i(\lambda)\phi_j(\lambda)\phi_k(\lambda)}{P'(\lambda)}
d\lambda
\label{resf}
\ee
In accordance with (\ref{defC}),
\be
G'(\lambda) = g^{(m)}\phi_m(\lambda)\ \ \ \ \ \ \
G_{ij} = g^{(m)}{\cal F}_{ijm}
\ee
The last ingredient is the expression of {\it flat} moduli
$a_i$ in terms of the polynomial $P(\lambda)$ \cite{KriW}:
\be
a_i = -{N\over i(N-i)}\res \left( P^{i\over N}dG \right),\ \
N = {\rm ord}(P)
\ee
These formulas (already used above in the proof of the existence of the WDVV
equations in the case of pure gluodynamics) have a straightforward
generalization to the
case of polynomials of several variables, $\phi_i(\vec\lambda)
= \phi_i(\lambda_1,\ldots,\lambda_n)$:
\be
\phi_i(\vec\lambda)\phi_j(\vec\lambda) =
C_{ij}^k\phi_k(\vec\lambda)Q(\vec\lambda) \ {\rm mod} \ \left(
\frac{\partial P}{\partial\lambda_1},\ldots,\frac{\partial P}
{\partial \lambda_n}\right),
\ee
and
\be
{\cal F}_{ijk} = \stackreb{dP = 0}{\res}\
\frac{\phi_i(\vec\lambda)\phi_j(\vec\lambda)\phi_k(\vec\lambda)}
{\prod_{\alpha=1}^n \frac{\partial P}{\partial \lambda_\alpha}}
d\lambda_1\ldots d\lambda_n
\label{resf1}
\ee
The algebra (\ref{algp}) is always associative, since
$dP = \sum_{\alpha=1}^n \frac{\partial P}{\partial \lambda_\alpha}
d\lambda_\alpha$ is always an ideal in the space of polynomials.
Moreover, one can even take a factor over generic ideal in the
space of polynomials, $p_1(\vec\lambda) = \ldots = p_n(\vec\lambda) = 0$,
where polynomials $p_\alpha$ need to be co-prime, but do not need to
be derivatives of a single $P(\vec\lambda)$. In this subsection we will
discuss in detail the algebraic structure underlying the associativity
equations which is more or less natural generalization of the (factorized)
polynomial ring structure.

The proof of the WDVV equations in the case of pure \N2 gluodynamics
presented above does not differ too much from the consideration in the
beginning of this subsection except for the substitution of
polynomials (functions on a Riemann sphere) holomorphic
1-differentials on Riemann surfaces (complex curves). They always form
a family of closed algebras, parametrized by a triple of
holomorphic differentials $dG,d{\cal W},d\Lambda$. However, these algebras
are {\em not} rings (since the product of two 1-differentials is already
a 2-differential), thus they do not give rise immediately to
associative algebra after factorization over an ideal. Still, associativity
is preserved for many important cases -- in particular for the hyperelliptic
curves.

The algebraic construction proposed in \cite{MMM4} is interesting because it
should possess direct generalizations to higher complex
dimensions (from holomorphic 1-forms on complex curves to forms on
complex manifolds), what physically means that one can pass from the
WDVV equations on the Seiberg-Witten prepotentials
to (hypothetical) universal
equations for the prepotentials in string models.

Hence, imagine that
in some context the following statements are true:
\begin{enumerate}
\item The holomorphic
\footnote{Since curves with punctures and the corresponding
meromorphic differentials
can be obtained by degeneration of smooth curves of higher genera we do not
make any distinction between punctured and smooth curves below. We remind that
the holomorphic 1-differentials can have at most simple poles at the punctures
while quadratic differentials can have certain double poles etc.}
1-differentials
on the complex curve $\Sigma $ of genus $g$ form a {\it closed} algebra,
\be
d\omega_i(\lambda)d\omega_j(\lambda) = C_{ij}^kd\omega_k(\lambda)
dG(\lambda) + D_{ij}^kd\omega_k(\lambda)d{\cal W}(\lambda)
+ E_{ij}^kd\omega_k(\lambda)d\Lambda(\lambda) = \nn \\ =
C_{ij}^kd\omega_k(\lambda)dG(\lambda) \ {\rm mod} \ (d{\cal W},d\Lambda),
\label{algdiff}
\ee
where $d\omega_i(\lambda)$, $i=1,\ldots, g$, form a complete basis in the
linear space $\Omega^1$ (of holomorphic 1-forms),
$dG$, $d{\cal W}$ and $d\Lambda$ are fixed elements of $\Omega^1$, e.g.
$dG(\lambda) = \sum_{m=1}^g  \eta^{(m)}d\omega_m$.
\item The factor of this algebra over the ``ideal'' $d{\cal W}\oplus d\Lambda$
is associative,
\be
C_i C_j = C_jC_i \ \ \forall  i,j \ {\rm at\ fixed}\
dG, d{\cal W}, d\Lambda
\label{comCdiff}
\ee
(remind that $(C_i)^k_j \equiv C_{ij}^k$).
\item The residue formula holds,
\be
\frac{\partial {\cal F}}{\partial a_i\partial_j\partial a_k}
= \stackreb{d{\cal W} = 0}{\res } \frac{d\omega_id\omega_jd\omega_k}
{d{\cal W}d\Lambda} =
-\stackreb{d\Lambda = 0}{\res} \frac{d\omega_id\omega_jd\omega_k}
{d{\cal W}d\Lambda}
\label{resfdiff}
\ee
\item There exists a non-degenerate linear combination of matrices
${\cal F}_i$.
\end{enumerate}
These statements imply the WDVV eqs (\ref{FFF}) for the prepotential
${\cal F}(a_i)$. Indeed, the
substitution of (\ref{algdiff}) into (\ref{resfdiff}) gives
\be
{\cal F}_{ijk} = C^m_{ij}G_{mk},
\label{int1}
\ee
where
\be
G_{mk} = \stackreb{d{\cal W} = 0}{\res} \frac{dG d\omega_md\omega_k}
{d{\cal W}d\Lambda} = \eta^{(l)}{\cal F}_{lmk},
\ee
and the terms with $d{\cal W}$ and $d\Lambda$ in (\ref{algdiff}) drop
out from ${\cal F}_{ijk}$ because they cancel $d\Lambda$ or $d{\cal W}$
in the denominator in (\ref{resfdiff}).
Eq.(\ref{int1}) can be now substituted into (\ref{comCdiff}) to
provide WDVV eqs in the form
\be
{\cal F}_i G^{-1} {\cal F}_j = {\cal F}_j G^{-1} {\cal F}_i,
\ \ \  G =  \eta^{(m)}{\cal F}_m \ \ \ \forall\  \left\{\eta^{(m)}\right\}
\label{WDVVdiff}
\ee
where at least one invertible metric $G$ exists by requirement (4).

Existence of the multiplication algebra (\ref{algdiff}) is rather
natural feature of complex curves. Indeed, there are $g$
holomorphic 1-differentials on the complex curve of genus $g$.
However, their products $d\omega_id\omega_j$ are not linearly
independent: they belong to the $3g-3$-dimensional space
$\Omega^2$ of the holomorphic quadratic differentials. Given
three holomorphic 1-differentials $dG$, $d{\cal W}$, $d\Lambda$, one can
make an identification
\be
\Omega^1\cdot\Omega^1 \in \Omega^2 \cong \Omega^1\cdot
(dG \oplus d{\cal W} \oplus d\Lambda)
\ee
which in particular basis is exactly (\ref{algdiff}).
For given $i,j$ there are $3g$ adjustment parameters $C_{ij}^k$,
$D_{ij}^k$ and $E_{ij}^k$ at the r.h.s. of (\ref{algdiff}), with
3 "zero modes" -- in the directions $dGd{\cal W}$, $dGd\Lambda$
and $d{\cal W}d\Lambda$ (i.e. one can add $d{\cal W}$ to $C_{ij}^kd\omega_k$ and
simultaneously subtract $dG$ from $D_{ij}^kd\omega_k$). Thus we get
exactly $3g-3$ parameters to match the l.h.s. of
(\ref{algdiff}) -- this makes decomposition (\ref{algdiff}) existing
and unique.

Thus we found that the existence of the {\it closed} algebra
(\ref{algdiff}) is a general feature, in particular it does not make any
restrictions on the choice of Riemann surfaces.
However, this algebra is not a ring: it maps
$(\Omega^1)^{\otimes 2}$ into {\it another} space:
$\Omega^1\otimes\Omega^1 \rightarrow
\Omega^2 \neq \Omega^1$. Thus, its factor over the condition
$d{\cal W} = d\Lambda = 0$ is not guaranteed to have all properties of the ring.
In particular the factor-algebra
\be
d\omega_i\circ d\omega_j = C_{ij}^kd\omega_k
\label{facalgdiff}
\ee
does not need to be associative, i.e. the matrices
$C$ alone (neglecting $D$ and $E$) do not necessarily commute.

However, the associativity would follow if the expansion of $\Omega^3$
(the space of the holomorphic 3-differentials containing
the result of triple multiplication $\Omega^1\cdot\Omega^1\cdot\Omega^1$),
\be
\Omega^3 = \Omega^1\cdot dG\cdot dG \oplus \Omega^2\cdot d{\cal W} \oplus
\Omega^2\cdot d\Lambda
\label{ass}
\ee
is unique. Then it is obvious that
\be
0 = (d\omega_id\omega_j)d\omega_k - d\omega_i(d\omega_jd\omega_k) =
\left( C_{ij}^lC_{lk}^m - C_{il}^mC_{jk}^l\right)d\omega_m dG^2
\ {\rm mod} (d{\cal W},d\Lambda)
\label{ass2}
\ee
would imply $[C_i,C_k] = 0$. However, the dimension of
$\Omega^3$ is $5g-5$, while the number of adjustment parameters
at the r.h.s. of (\ref{ass}) is $g + 2(3g-3) = 7g-6$, modulo only
$g+2$ zero modes (lying in $\Omega^1\cdot d{\cal W}d\Lambda$,
$\Omega^1\cdot d{\cal W}dG^2$ and $\Omega^1\cdot d\Lambda dG^2$). For $g>3$ there
is no match:
$5g-5 < 6g-8$, the expansion (\ref{ass}) is not unique, and
associativity can (and does)
\footnote{See \cite{MMM3} for an explicit example of {\it non}-associativity
(actually, this happens in the important Calogero model).}
break down unless there is some special reason for it to survive.

This special reason can exist if the curve $\Sigma $ has specific symmetries.
The most important example is the set of curves
with an involution $\sigma:\ \Sigma  \rightarrow \Sigma $,
$\sigma^2=1$, such that all
$\sigma(d\omega_i) = -d\omega_i$, while $\sigma(d{\cal W}) = -d{\cal W}$,
$\sigma(d\Lambda) = +d\Lambda$. To have $d\Lambda$ different from
all $d\omega_i$ one should actually take it away from $\Omega^1$,
e.g. allow it to be meromorphic.

The most well-known particular example of such curves is the family of
hyperelliptic curves described by the equation
\be
Y^2 = {\rm Pol}_{2g+2}(\lambda),
\ee
and the involution is $\sigma: (Y,\lambda )
\rightarrow (-Y, \lambda )$. The space of holomorphic differentials
is $\Omega^1 = {\rm Span}\left\{\frac{\lambda^\alpha d\lambda}{Y(\lambda)}
\right\}$,
$\alpha = 0,\ldots,g-1$.  This space is odd under $\sigma$,
$\sigma(\Omega^1) = -\Omega^1$, and an example of the (meromorphic)
1-differential which is {\it even} is
\be
d\Lambda = \lambda^rd\lambda,
\ee
$\sigma(d\Lambda) = +d\Lambda$. We will assume that $dG$ and $d{\cal W}$ still belong
to $\Omega^1$ and thus are $\sigma$-odd. In the case of hyperelliptic
curves with punctures, $\Omega^1$ can include also $\sigma$-even
holomorphic 1-differentials (like $\frac{d\lambda}{(\lambda - \alpha_1)
(\lambda - \alpha_2)}$ or just $d\Lambda$), in such cases we consider
the algebra (\ref{algdiff}) of the $\sigma$-odd holomorphic differentials
$\Omega^1_-$, and assume that $d\omega_i$, $dG$ and $d{\cal W}$ belong to
$\Omega^1_-$, while $d\Lambda \in \Omega^1_+$.

The spaces $\Omega^2$ and $\Omega^3$ also split into $\sigma$-even
and $\sigma$-odd parts: $\Omega^2 = \Omega^2_+ \oplus \Omega^2_-$
and $\Omega^3 = \Omega^3_+ \oplus \Omega^3_-$. Multiplication
algebra maps $\Omega^1_-$ into $\Omega^2_+$ and further into
$\Omega^3_-$, which have dimensions  $2g-1+2n$ and $3g-2+3n$
respectively. Here $n$ enumerates the punctures, where holomorphic
1-differentials are allowed to have simple poles, while quadratic
and the cubic ones
have at most second- and third-order poles respectively. For our
purposes we assume that punctures on the hyperelliptic curves
enter in pairs: every
puncture is accompanied by its $\sigma$-image.
Parameter $n$ is the number of these {\it pairs},
and the dimension of $\Omega^1_-$ is $g+n$.

Obviously, if all the $d\omega_i$ in (\ref{algdiff})
are from $\Omega^1_-$, then all $E^k_{ij} = 0$, i.e. we actually
deal with the decomposition
\be
\Omega^2_+ = \Omega^1_-\cdot dG + \Omega^1_-\cdot d{\cal W}
\label{2dec}
\ee
Parameter count now gives:
$2g-1 + 2n = 2(g+n) - 1$ where $-1$ is for the zero mode
$dGd{\cal W}$. Thus, the hyperelliptic reduction of the algebra (\ref{algdiff})
does exist.

Moreover, it is associative, as follows from consideration
of the decomposition
\be
\Omega^3_- = \Omega^1_-\cdot dG^2 + \Omega^2_+ \cdot d{\cal W}
\label{3dec}
\ee
Of crucial importance is that now there is no need to include
$d\Lambda$ in this decomposition, since it does not appear at the
r.h.s. of the algebra itself.
Parameter count is now:
$3g-2 + 3n = (g+n) + (2g-1+ 2n) -1$ (there is the unique zero mode
$d{\cal W}dG^2$). Thus, we see that this time decomposition (\ref{2dec})
is unique, and our algebra is indeed associative.

In fact, one could come to the same conclusions much easier just
noting that all elements of $\Omega^1_-$ are of the form
\be
d\omega_i = \frac{\phi_i(\lambda) d\lambda}{YQ(\lambda)},
\ee
where all $\phi_i(\lambda)$ are polynomials and $Q(\lambda)
= \prod_{\iota =1}^n(\lambda - m_{\iota})$ is
some new polynomial, which takes into account the possible
singularities at punctures $\left(m_{\iota },\pm Y(m_{\iota })\right)$.
Then our algebra is just the one of the polynomials $\phi_i(\lambda)$
and it is existing and associative just for the reasons discussed
before. The reasoning in this section can be easily
modified in the case when hyperelliptic curve possesses an extra involution.
The families of such curves appear in the Seiberg-Witten context for the
groups $SO(N)$ and $Sp(N)$: the extra involution in these cases is
$\rho :\lambda\rightarrow -\lambda$. Then one considers $\Omega^1_{--}$
instead of just $\Omega^1_-$ (see \cite{MMM3} for further details).

Let us return to the most general consideration of the residue formulas.
Consider an integrable model
with a Lax operator ${\cal L}(w)$, which is a $N\times N$
matrix-valued function (see sect.2 where several examples of such
models related to the non-perturbative effective gauge models are considered
in details: for them $N\equiv N_c$)
on a bare spectral curve $E$, $w \in E$, which is
usually torus or sphere.
Then one can introduce a family of complex curves,
defined by the spectral equation (cf. with eqs (\ref{SpeC}), (\ref{fscCal}))
\be
\det({\cal L}(w)-\lambda) = 0
\ee
The family is parametrized by the {\it moduli} that in this
context are values of the $N$ Hamiltonians of the system
(since Hamiltonians commute with each other, these are actually
$c$-numbers). We obtain this family in a peculiar parametrization,
which represents the {\it full} spectral curves $\Sigma $
as the ramified $N$-sheet coverings  over the  {\it bare} curve
$E$,
\be
{\cal P}(\lambda; w) = 0,
\label{curveq}
\ee
where ${\cal P}$ is a polynomial of degree $N$ in $\lambda$.

Integrable system is defined by a ``generating'' 1-form
$dS = \Lambda d{\cal W}$, which possesses the property:
\be
\frac{\partial dS}{\partial {\rm moduli}} \in \Omega^1,
\label{holreq}
\ee
i.e. every variation of $dS$ with the change of moduli is
a holomorphic differential on $\Sigma $ (normally, even if
differential is holomorphic, its moduli-derivative is not).

This structure allows one to define the (subfamily of) holomorphic
differentials in a rather explicit form. Let $s_I$ denote some (specific)
coordinates on moduli space ${\cal M}$. Then
\be
\frac{\partial dS}{\partial s_I} \cong \frac{\partial\Lambda}
{\partial s_I} d{\cal W} = -\frac{\partial {\cal P}}{\partial s_I}
\frac{d{\cal W}}{{\cal P}'} \equiv dv_I,
\label{defdv}
\ee
and $dv_I$ provide a set (in general not a canonical set) of holomorphic
differentials on $\Sigma $.
The set of $dv_I$ is not necessarily the same as $\Omega^1_-$,
it can be either a subspace of $\Omega^1_-$ or some $dv_I$ can be linearly
dependent. It is a special requirement (standard in the context of
integrable theories) that the differentials $dv_I$'s form a complete
basis in $\Omega^1_-$ (or in $\Omega^1_{--}$).

The prepotential ${\cal F}(a_I)$ is defined by standard formulas
(\ref{aper}), (\ref{adper}), (\ref{flogt}) and (\ref{defF}). The
definition implies in general that the cycles $A_I$
include $A_\iota$'s going around the punctures.
The conjugate contours $B_\iota$ ending in the singularities
of $dS$.

The self-consistency of the definition (\ref{defF}) of ${\cal F}$,
i.e. the symmetricity of the {\it period matrix}
$\frac{\partial^2 F}{\partial a_I\partial a_J}$ is guaranteed
by the following reasoning.
Let us differentiate equations (\ref{defF}) with respect to
moduli $s_K$ and use (\ref{defdv}). Then we get:
\be
\int_{B_I} dv_K = \sum_J T_{IJ}\oint_{A_J} dv_K.
\ee
where the second derivative
\be
\frac{\partial^2 F}{\partial a_I\partial a_J} = T_{IJ}
\ee
is the period matrix of the (punctured) Riemann surface
$\Sigma $.
As any period matrix, it is symmetric
\be
\sum_{IJ} (T_{IJ} - T_{JI}) \oint_{A_I}dv_K\oint_{A_J}dv_L =
\sum_I\left(\oint_{A_I}dv_K\int_{B_I}dv_L -
\int_{B_I}dv_K\oint_{A_I}dv_L\right) =  \nn \\ =
{\res}\left( v_K dv_L \right) = 0
\ee
Note also that the holomorphic differentials
associated with the {\it flat} moduli $a_I$ are the {\it canonical}
$d\omega_I$ such that $\oint_{A_I}d\omega_J = \delta_{IJ}$
and $\oint_{B_I}d\omega_J = T_{IJ}$.

In order to derive the residue formula one should now consider
the moduli derivatives of the period matrix.
It is easy to get:
\be
\sum_{IJ} \frac{\partial T_{IJ}}{\partial s_M}
\oint_{A_I}dv_K\oint_{A_J}dv_L =
\sum_I\left(\oint_{A_I}dv_K\int_{B_I}\frac{\partial dv_L}{\partial s_M} -
\int_{B_I}dv_K\oint_{A_I}\frac{\partial dv_L}{\partial s_M}\right) =
\nn \\ =
\res \left( v_K \frac{\partial dv_L}{\partial s_M}\right)
\label{prom1}
\ee
The r.h.s. is non-vanishing, since differentiation w.r.t. moduli
produces new singularities. From (\ref{defdv})
\be
-\frac{\partial dv_L}{\partial s_M} =
\frac{\partial^2{\cal P}}{\partial s_L\partial s_M}
\frac{d{\cal W}}{{\cal P}'} +
\left(\frac{\partial{\cal P}}{\partial s_L}\right)'
\left(-\frac{\partial{\cal P}}{\partial s_M}\right)
\frac{d{\cal W}}{{\cal P}'} -
\frac{\partial{\cal P}}{\partial s_L}
\frac{\partial{\cal P}'}{\partial s_M}
\frac{d{\cal W}}{({\cal P}')^2} +
\frac{\partial{\cal P}}{\partial s_L}
\frac{\partial{\cal P}}{\partial s_M}
{{\cal P}''d{\cal W}\over ({\cal P}')^3}
= \nn \\ =
\left[
\left(\frac{ {\partial{\cal P}/\partial s_L}
 {\partial{\cal P}/\partial s_M} }{ {\cal P}'}  \right)' +
\frac{\partial^2{\cal P}}{\partial s_L\partial s_M} \right]
\frac{d{\cal W}}{{\cal P}'}
\ee
and new singularities (second order poles)
are at zeroes of ${\cal P}'$ (i.e. at those of $d{\cal W}$). Note
that the contributions from the singularities of
$\partial{\cal P}/\partial s_L$, if any, are already taken into
account in the l.h.s. of (\ref{prom1}).
Picking up the coefficient at the leading singularity, we obtain:
\be
{\res} v_K \frac{\partial dv_L}{\partial s_M} = -
\stackreb{d{\cal W} = 0}{\res}
\frac{\partial{\cal P}}{\partial s_K}
\frac{\partial{\cal P}}{\partial s_L}
\frac{\partial{\cal P}}{\partial s_M}
\frac{d{\cal W}^2}{({\cal P}')^3 d\Lambda} =
\stackreb{d{\cal W} = 0}{\res}
\frac{dv_Kdv_Ldv_M}{d{\cal W}d\Lambda}
\ee
The integrals at the l.h.s. of (\ref{prom1}) serve to convert
the differentials $dv_I$ into canonical $d\omega _I$. The same matrix
$\oint_{A_I} dv_J$ relates the derivative w.r.t. the moduli $s_I$
and the periods $a_I$. Putting all together we obtain
(see also \cite{KriW}):
\be
\frac{\partial T_{IJ}}{\partial s^K} =
\stackreb{d{\cal W} = 0}{\res}
\frac{d\omega_Id\omega_Jdv_K}{d{\cal W} d\Lambda} \nn \\
\frac{\partial^3 F}{\partial a^I\partial a^J\partial a^K} =
\frac{\partial T_{IJ}}{\partial a^K} =
\stackreb{d{\cal W} = 0}{\res}
\frac{d\omega_Id\omega_Jd\omega_K}{d{\cal W} d\Lambda}
\label{resfor}
\ee
Note that these formulas essentially depend only on the symplectic
structure $d{\cal W} \wedge d\Lambda$: e.g. if one makes an infinitesimal shift of
$d{\cal W}$ by $d\Lambda$, then $(d{\cal W}d\Lambda)^{-1}$ is shifted by
$-(d{\cal W})^{-2}$, i.e. the shift does not contain poles at $d\Lambda = 0$
and thus does not contribute to the residue formula.
Let us note finally, that the above considerations and the residue formulas
should be applied literally to the systems described by the holomorphic
differentials. In case of curves with the marked points (and, correspondingly,
the differentials with poles in these points) the presented above formulas
needs some more accurate extra definition, which could be found in \cite{MMM4}.

\subsection{Perturbative example}

{\bf The explicit example of the solution to the associativity equations
in the framework of the effective Seiberg-Witten theory}.
This example corresponds to the perturbative part
of the Seiberg-\-Witten prepotential for ${\cal N}=2$ SUSY gluodynamics
which itself satisfies equations (\ref{FFF}).
Since the perturbative contribution is non-transcendental, the
calculation can be performed in explicit form:
\be
{\cal F}_{pert} \equiv {\cal F}(a_i) =
\left.\frac{1}{2}\sum_{\stackrel{m<n}{m,n=1}}^N
(A_m-A_n)^2\log(A_m-A_n)\right|_{\sum_m A_m = 0} = \nn \\
= \frac{1}{2}\sum_{\stackrel{i<j}{i,j=1}}^{N-1}
(a_i-a_j)^2\log(a_i-a_j) +
\frac{1}{2}\sum_{i=1}^{N-1}a_i^2\log a_i
\label{pertF}
\ee
Here we took $a_i = A_i - A_N$ -- one of the many
possible choices of independent variables, which differ by
linear transformations. According to (\ref{F-1})
the system (\ref{FFF}) is covariant under such changes.

Formula (\ref{pertF}) means that in $4d$ SUSY YM theory the perturbative
contribution to the effective action has the structure
\be\label{masses}
{\cal F}_{pert} =  \frac{1}{4}\sum_{\rm masses} ({\rm mass})^2 \log ({\rm mass})
\ee
where in (\ref{pertF}) all masses are generated by the v.e.v.'s of the
Higgs field by spontaneous breaking mechanism. The formula (\ref{masses})
comes from the requirement that the effective charge
\be\label{efcharge}
\delta ^2{\cal F}_{pert} \sim \sum_{\rm masses} \log ({\rm mass})
\ee
is one-loop and pure logarithmic.

Let us introduce the notation $a_{ij} = a_i - a_j$. The matrix
\be
\{({\cal F}_1)_{mn}\} = \left\{\frac{\partial^3 {\cal F}}{\partial a_1
\partial a_m\partial a_n} \right\} = \nn \\ =
\left(\begin{array}{ccccc}
\frac{1}{a_1} +\sum_{l\neq 1} \frac{1}{a_{1l}} &
-\frac{1}{a_{12}} & -\frac{1}{a_{13}} & -\frac{1}{a_{14}} & \\
-\frac{1}{a_{12}}& +\frac{1}{a_{12}} & 0 & 0 & \\
-\frac{1}{a_{13}}& 0 & +\frac{1}{a_{13}}& 0 &\ldots \\
-\frac{1}{a_{14}}& 0 & 0 &+\frac{1}{a_{14}} & \\
&&\ldots && \end{array}\right)
\ee
i.e.,
\be\label{f}
\{({\cal F}_i)_{mn}\} = \frac{\delta_{mn}(1-\delta_{mi})(1-\delta_{ni})}
{a_{im}} - \frac{\delta_{mi}(1-\delta_{ni})}{a_{in}}
- \frac{\delta_{ni}(1-\delta_{mi})}{a_{im}} + \nn \\ +
\left(\frac{1}{a_i} + \sum_{l\neq i}\frac{1}{a_{ik}}\right)
\delta_{mi}\delta_{ni}
\ee
The inverse matrix
\be\label{f-1}
\{({\cal F}_k^{-1})_{mn}\} = a_k + \delta_{mn}a_{km}(1-\delta_{mk}),
\ee
for example
\be
\{({\cal F}_1^{-1})_{mn}\} = a_1\left(\begin{array}{cccc}
1 & 1 & 1 & . \\ 1 & 1 & 1 & . \\ 1 & 1 & 1 & . \\
&\ldots & & \end{array}\right) +
\left(\begin{array}{cccc}
0 & 0 & 0 & . \\
0 & a_{12} & 0 & . \\
0 & 0 & a_{13} & . \\
& \ldots & & \end{array}\right)
\ee
As the simplest example let us consider the case $N=4$.
We already know that for $N=4$ it is
sufficient to check only one of the eqs.(\ref{FFF}),
all the others follow automatically. We take $k=1$. Then,
\be
{\cal F}_1=\left(
\begin{array}{ccc}
{1\over a_1}+{1\over a_{12}}+{1\over a_{13}}&-{1\over a_{12}}&-{1\over
a_{13}} \\-{1\over a_{12}}&{1\over a_{12}}&0\\-{1\over a_{13}}&0&{1\over
a_{13}}
\end{array}\right)\ \ \ {\cal F}^{-1}_2=\left(
\begin{array}{ccc}
a_2+a_{21}&a_2&a_2\\a_2&a_2&a_2\\a_2&a_2&a_2+a_{23}
\end{array}\right)\\ {\cal F}_3=\left(
\begin{array}{ccc}
{1\over a_{31}}&0&-{1\over a_{31}}\\
0&{1\over a_{32}}&-{1\over a_{32}}\\
-{1\over a_{31}}&-{1\over a_{32}}&{1\over a_3}+{1\over a_{31}}+{1\over
a_{32}}
\end{array}\right)
\ee
and, say,
\be
{\cal F}_1{\cal F}^{-1}_2{\cal F}_3=\left(
\begin{array}{ccc}
\star&-{1\over a_{31}}& \Delta + {a_{21}+a_{23}\over a_{13}^2}\\
-{1\over a_{13}}&\star&{1\over a_{13}}\\
{a_{21}+a_{23}\over a_{13}^2}&{1\over a_{13}}&\star
\end{array}\right)
\ee
where we do not write down manifestly the diagonal terms since, to check
(\ref{FFF}), one only needs to prove the symmetricity of the matrix. This is
really the case, since
\be
\Delta\equiv
{a_2\over a_1a_3}-{a_{21}\over a_1a_{31}}-{a_{23}\over a_3a_{13}} =0
\ee
Only at this stage we use manifestly that $a_{ij}=a_i-a_j$.

Now let us prove (\ref{FFF}) for the general case. We check the equation
for the inverse matrices. Namely, using formulas (\ref{f})-(\ref{f-1}), one
obtains
\be\label{long}
({\cal F}_i^{-1}{\cal F}_j{\cal F}_k^{-1})_{\alpha\beta}=
\\
={a_ia_k\over a_j}+
\delta_{\alpha\beta}(1-\delta_{i\alpha})(1-\delta_{k\alpha})
(1-\delta_{j\alpha}){a_{i\alpha}a_{k\beta}\over a_{j\beta}}+
\delta_{j\alpha}\delta_{j\beta}(1-\delta_{i\alpha})(1-\delta_{k\beta})
\left({1\over a_j}+\sum_{n\ne j}{1\over a_{jn}}\right)+\\
+\delta_{j\alpha}(1-\delta_{i\alpha})a_{i\alpha}\left(
{a_k\over a_j}-{a_{k\beta}\over a_{j\beta}}(1-\delta_{k\beta})
(1-\delta_{j\beta})\right)+\delta_{j\beta}(1-\delta_{k\beta})\left(
{a_i\over a_j}-{a_{i\alpha}\over a_{j\alpha}}(1-\delta_{i\alpha})
(1-\delta_{j\alpha})\right)=\\
={a_ia_k\over a_j}+
\delta_{\alpha\beta}(1-\delta_{i\alpha}-\delta_{k\alpha}
-\delta_{j\alpha}){a_{i\alpha}a_{k\beta}\over a_{j\beta}}+
\delta_{j\alpha}\delta_{j\beta}\left({1\over a_j}+
\sum_{n\ne j}{1\over a_{jn}}\right)+\\
+\delta_{j\alpha}a_{i\alpha}\left(
{a_k\over a_j}-{a_{k\beta}\over a_{j\beta}}(1-\delta_{k\beta}
-\delta_{j\beta})\right)+\delta_{j\beta}\left(
{a_i\over a_j}-{a_{i\alpha}\over a_{j\alpha}}(1-\delta_{i\alpha}
-\delta_{j\alpha})\right)
\ee
where we used that $i\ne j\ne k$. The first three terms are evidently
symmetric with respect to interchanging $\alpha\leftrightarrow\beta$. In
order to prove the symmetricity of the last two terms, we need to use the
identities ${a_k\over a_j}-{a_{k\beta}\over a_{j\beta}}={a_{\beta}a_{jk}\over
a_ja_{j\beta}}\stackrel{k=\beta}{\to}{a_k\over a_j}$,
${a_i\over a_j}-{a_{i\alpha}\over a_{j\alpha}}=
{a_{\alpha}a_{ji}\over a_ja_{j\alpha}}\stackrel{i=\alpha}{\to}{a_i\over
a_j}$. Then, one gets
\be
\hbox{the last line of (\ref{long})}=
\delta_{j\alpha}(1-\delta_{j\beta})
{a_{ij}a_{jk}\over a_j}{a_{\beta}\over a_{j\beta}} +
\delta_{j\beta}(1-\delta_{j\alpha})
{a_{ij}a_{jk}\over a_j}{a_{\alpha}\over a_{j\alpha}} +
\delta_{j\alpha}\delta_{j\beta}{a_ka_{i\alpha}+a_ia_{k\beta}\over a_j}
\ee
It is interesting to note that
in the particular example (\ref{pertF}),
all the metrics $\eta^{(k)}$ are flat. Moreover,
it is easy to find the explicit flat coordinates:
\be
\eta^{(k)} = \eta^{(k)}_{ij}da^ida^j =
{\cal F}_{ijk}da^ida_j = da_ida_j\partial^2_{ij}(\partial_k {\cal F}) = \nn \\ =
\frac{da_k^2}{a_k} + \sum_{l\neq k}\frac{da_{kl}^2}{a_{kl}} = 4\left(
(d\sqrt{a_k})^2 + \sum_{l\neq k}(d\sqrt{a_{kl}})^2\right).
\ee
The explicit form of non-perturbative (instantonic etc) contributions
to the prepotential are more complicated. For several examples their
computation from the exact formulas {\em a la} Seiberg-Witten can be found
in \cite{MMM3} and the discussion of consistency with standard quantum field
theory methods can be found in \cite{khoze}.

{\bf Holomorphic differentials on a punctured sphere}. Let us show now that
the perturbative example corresponds to {\em rational} degeneration of
the spectral curve, namely to the Riemann sphere with some punctures at the
points $\lambda_i$, $i=1,\ldots,N$, so that
the canonical basis in the space $\Omega^1$ can be chosen as:
\be
d\omega_i = \frac{(\lambda_i-\lambda_N)d\lambda}{(\lambda - \lambda_i)
(\lambda - \lambda_N)}, \ \ \ i=1,\ldots,N-1
\ee
We assumed that the $A_i$ cycles wrap around the points $\lambda_i$,
while their conjugated $B_i$ connect $\lambda_i$ with the reference
puncture $\lambda_N$. Multiplication algebra of $d\omega_i$'s
is defined modulo
\be
d{\cal W} = d\log P_N(\lambda) = \frac{dP_N(\lambda)}{P_N(\lambda)},
\ee
$P_N(\lambda) = \prod_{i=1}^N (\lambda - \lambda_i)$, and
it is obviously associative.

The periods $a_i$ depend on the choice of the generating
differential $dS = \Lambda d{\cal W}$. There are two essentially different
choices $\Lambda = \lambda$ \cite{GKMMM} and $\Lambda = \log\lambda$
\cite{Nekrasov}, i.e.
\be
dS^{(4)} = \lambda \ d\log P_N(\lambda) \ \ \ {\rm and} \ \ \
dS^{(5)} = \log\lambda\  d\log P_N(\lambda)
\ee
In order to fulfill the requirement (\ref{holreq}) one should
assume that $\sum_{i=1}^N \lambda_i = 0$ in the case of
$dS^{(4)}$, while $\prod_{i=1}^N \lambda_i = 1$ in the case
of $dS^{(5)}$. Since $A_i$ cycle just wraps around the point $\lambda
=\lambda_i$, the $A_i$-periods of such $dS$ are
\be
a_i^{(4)}=\oint_{\lambda_i}dS^{(4)}=\lambda_i,\\
a_i^{(5)}=\oint_{\lambda_i}dS^{(5)}=\log\lambda_i
\ee
The corresponding residue formulas are
\be
{\cal F}^{(4)}_{ijk} = \sum_{m=1}^N \stackreb{\lambda_m}{\res}
\frac{d\omega_id\omega_jd\omega_k}{d\lambda\ d\log P_N}, \nn \\
{\cal F}^{(5)}_{ijk} = \sum_{m=1}^N \stackreb{\lambda_m}{\res}
\lambda\frac{d\omega_id\omega_jd\omega_k}{d\lambda\ d\log P_N},
\ \ \ i,j,k=1,\ldots,N-1
\ee
and they both provide solutions to the WDVV equations. The prepotentials
are (\ref{pertF}) and \cite{MMM3,MMM4}:
\be\label{F5d}
{\cal F}^{(5)}(a_i) =\sum_{1\leq i < j \leq N}
\widetilde Li_3\left(e^{a_i-a_j}\right)-
{N\over 2}\sum_{1\leq i < j < k \leq N} a_i a_j a_k, \ \ \
\sum_{i=1}^N a_i = 0,\\
\partial^2_x \widetilde Li_3\left(e^x\right)\equiv \log 2\sinh x,\ \ \
\widetilde Li_3\left(e^x\right)={1\over 6} x^3 -{1\over 4}
Li_3\left(e^{-2x}\right)
\ee
and describe the perturbative
limit of the $N=2$ supersymmetric $SU(N)$ gauge models in $4d$
and $5d$ \cite{Nekrasov} respectively. Note only that the expression (\ref{F5d})
differs from that of \cite{Nekrasov} on cubic in periods ${\bf a}$
terms, whose presence is {\em necessary} for the prepotential to
satisfy the WDVV equations.

If the punctures $\lambda_i$  are not all independent,
the same formulas provide solutions to the WDVV equations,
associated with the other simple groups: $SO(N)$,
$Sp(N)$, $F_4$ and $E_{6,7,8}$ ($G_2$ does not have enough
moduli to provide non-trivial solutions to the WDVV eqs).
If $P_N$ is substituted by
\be
P_N \rightarrow \frac{P_N}{Q_{N_f}^{1/2}} = \frac{\prod_{i=1}^N
(\lambda - \lambda_i)}{\prod_{\iota = 1}^{N_f} (\lambda - m_\iota)^{1/2}},
\ee
one gets solutions, interpreted as (perturbative limits of) the gauge models
with matter supermultiplets in the first fundamental representation.
Inclusion of matter in other representations seems to
destroy the WDVV equations, at least, generically; note that such models do
not arise in a natural way from string compactifications, and there
are no known curves associated with them in the Seiberg-Witten theory
(see \cite{MMM3} for details).

{\bf Holomorphic differentials on hyperelliptic curves}.
{\it Non-perturbative} deformations of the above prepotentials
arise when the punctures on Riemann sphere are blown up to form
handles of the hyperelliptic curve:
\be
W + \frac{1}{W} = 2\frac{P_N(\lambda)}{Q(\lambda)^{1/2}_{N_f}}, \nn \\
W - \frac{1}{W} = 2\frac{Y(\lambda)}{Q(\lambda)^{1/2}_{N_f}}, \nn \\
Y^2(\lambda) = P_N^2(\lambda) - Q_{N_f}(\lambda)
\ee
These curves, together with the corresponding differentials $dS$
\be
dS^{(4)} = \lambda\frac{dW}{W}, \ \ \
dS^{(5)} = \log\lambda \frac{dW}{W},
\ee
(i.e. $d{\cal W} = \frac{dW}{W}$ and $d\Lambda^{(4)} = d\lambda$, $d\Lambda^{(5)}
= \frac{d\lambda}{\lambda}$) are implied by integrable models of the
Toda-chain family \cite{GKMMM,MartW1,Tak,Nekrasov}.
Together with the residue formula (\ref{resfdiff}) these provide the
nonperturbetive solution to the WDVV equations.

For example the proof of subsect.3.2 can be almost literally transfered to
the case of the
relativistic Toda chain system corresponding to the $5d$ $N=2$ SUSY pure gauge
model with one compactified dimension \cite{Nekrasov}. The main reason is
that in the case of relativistic Toda chain the spectral curve is a minor
modification of (\ref{specTC}) having the form
\be\label{specRTC}
w + {1\over w} = \left(\zeta\lambda\right)^{-N_c/2}P(\lambda ),
\ee
which can be again rewritten as a {\em hyperelliptic} curve in terms of the
new variable $Y\equiv \left(\zeta\lambda\right)^{N_c/2}\left(w-
{1\over w}\right)$
\be
Y^2 = P^2(\lambda ) - 4\zeta^{2N_c}\lambda^{N_c}
\ee
where $\lambda \equiv e^{2\xi}$, $\xi $ is the "true" spectral
parameter of the relativistic Toda chain and $\zeta$ is its coupling constant.

The difference with the $4d$ case (see subsect.3.2) is the following two
points. The first is that now $s_0\sim \prod e^{a_i}=1$ while
$s_{N_c-1}$ becomes an unfrozen moduli parameter.

The second essential new point is that the generating differential instead of
(\ref{dS}) in five-dimensional case is
\be\label{dSRTC}
dS = \xi{dw\over w} \sim \log\lambda {dw\over w}
\ee
so that
\be
dW_k = {\partial dS\over\partial s_k} \cong {\lambda^{k-1}d\lambda\over Y},
\ \ \ k=1,...,g
\ee
Despite the condition $s_0=1$, i.e. absence of the
corresponding module, this formula literally coincides
(because of the additional degree $\lambda$ in the
denominator) with formula (\ref{sigmadef}). Thus, the algebra of
differentials remains the same associative algebra,
the only difference being slightly
modified residue formula because of modifying the polynomial $P$ and,
therefore, the differential $d\omega$. Namely, the residue formula
acquires the following form
\be
F_{ijk} =
\stackreb{d\lambda=0}{{\res}} \frac{d\omega_id\omega_j
d\omega_k}{\left(\frac{d\lambda}{\lambda}\right)\left(\frac{dw}{w}\right)} =
\stackreb{d\lambda=0}{{\res}} \frac{d\omega_id\omega_j
d\omega_k}{\left(\frac{d\lambda}{\lambda}\right)\left(\frac{dP}{Y}\right)} =
\sum_{\alpha} \lambda_{\alpha}\frac{\hat\omega_i(\lambda_\alpha)\hat\omega_j
(\lambda_\alpha)\hat\omega_k(\lambda_\alpha)}{P'(\lambda_\alpha)
/\hat Y(\lambda_\alpha)}
\ee

{\bf Other examples}.
The very natural question is what happens with the WDVV equations
for Toda chain models, associated with the exceptional groups. The
problem is that the associated spectral curves are not hyperelliptic --
at least naively. Still they have enough symmetries to make our
general reasoning working, but this requires a special investigation.

The number of examples can be essentially increased by the study of
various integrable hierarchies, peculiar configurations of punctures
etc. In recent paper \cite{KriL} it was actually suggested that
-- at least in peculiar models -- $dS$ can be expressed through the
Baker-Akhiezer function: $dS = \Lambda d\log\Psi$.
Of more importance could crucially interesting lift to $6d$ or
elliptic models (in the same sense as five-dimensional theory
is a cyllindric or trigonometric $\lambda\rightarrow\log\lambda$
generalization of the four-dimensional theory) which
requires interpretation of $\lambda $ as a coordinate on
elliptic curve. It is not yet known, if this transition breaks down the
WDVV equations.

\section{Conclusion}

In these notes I have tried to present the main ingredients of
the theory of integrable systems which appeared recently to be rather useful
to understand the nonperturbative results in quantum strings and
supersymmetric gauge theories. The most exciting thing in this picture
is that there exists an effective description (by means of classical and
finite-dimensional integrable models)
of the theory which is quantum (infinite-dimensional!) field theory,
contains propagating (massless) particles and is {\em not} a quantum integrable
model at least in conventional sense.

Hypothetically, a generalization to realistic string models is straightforward
and related first of all with the prepotentials arising in the study of the
models related to the Calabi-Yau compactifications.
The steps described above can be in principle repeated
leading to the integrable models based on the {\it higher-dimensional
complex} manifolds (instead of $1_{\bf C}$-dimensional {\em curves} $\Sigma$).
Such
integrable systems are not investigated yet in detail (see however
\cite{Hi,Mark}) and are supposed to be much more complicated than the
well-known integrable systems of KP/Toda type.

One more direction (which was almost not considered in the text above) is
related to the study of effective theories on
(partially) compactified target-\-spaces \cite{SW3,Nekrasov}. Adding
one compactified dimension leads to appearance of the well-\-known
class of relativistic integrable models \cite{Ruj} and allows one to
interpret the divisor on a complex curve corresponding to a finite-\-gap
solution in terms of the nonlocal observables (loops). Thus, on one hand it
should clarify
the sense of arising integrable systems, while on the other hand it is
a step towards study of the prepotentials of string models (see for example
\cite{strpot}) where there are contributions having similiar (though more
complicated) structure.

In spite of all the problems it is easy to believe that for all the theories
where it is possible to make any statement about the nonperturbative and
exact quantities there exists something more than a summation of a
perturbation theory. The main idea advocated above and had been
checked
already in several examples is based on general belief that the realistic
theory should be selfconsistent and adjust automatically its
properties not to be ill-\-defined both at large and small distances.
It looks that an adequate nonperturbative language for the effective
formulation of consistent in this sense field and string theories can be
looked for among integrable systems.

\section{Acknowledgments}

I am deeply indebted to B.Dubrovin, V.Fainberg, V.Fock, A.Gorsky,
A.Losev, N.Nekrasov, A.Orlov, A.Rosly, I.Tyutin, B.Voronov and A.Zabrodin and
especially to I.Krichever, A.Levin, A.Mironov and A.Morozov for many
illuminating discussions. The work was in
part supported by RFFI grant 96-02-19085 and INTAS grant 93-0633.


\begin{thebibliography}{12}

\bibitem{GKMMM}
A.Gorsky, I.Krichever, A.Marshakov, A.Mironov and A.Morozov,
Phys.Lett. {\bf B355} (1995) 466; hepth/9505035.

\bibitem{MartW1}
E.Martinec and N.Warner, hepth/9509161.

\bibitem{Tak}
T.Nakatsu and K.Takasaki, hepth/9509162.

\bibitem{WiDo}
R.Donagi and E.Witten, hepth/9510101.

\bibitem{Mart}
E.Martinec, hepth/9510204.

\bibitem{GM}
A.Gorsky and A.Marshakov, hepth/9510224, Phys.Lett. {\bf B375} (1996) 127.

\bibitem{MartWa}
E.Martinec and N.Warner, hepth/9511052.

\bibitem{IM}
H.Itoyama and A.Morozov, hepth/9511126;hepth/9512161;hepth/9601168.

\bibitem{M4}
A.Marshakov, Mod.Phys.Lett. {\bf A11} (1996) 1169; hepth/9602005.

\bibitem{AN}
C.Ann and S.Nam, hepth/9603028.

\bibitem{Spin}
A.Gorsky, A.Marshakov, A.Mironov and A.Morozov, hepth/9603140,
Phys.Lett. {\bf B380} (1996) 75.

\bibitem{Spin2}
A.Gorsky, A.Marshakov, A.Mironov and A.Morozov, hepth/9604078, in
{\it Problems in Modern Theoretical Physics}, Dubna 1996, 44-62.

\bibitem{KriPho}
I.Krichever and D.Phong, hepth/9604199.

\bibitem{M5} A.Marshakov,
{\sl From Nonperturbative Supersymmetric Quantum Gauge Theories to
Integrable Systems}, preprint FIAN/TD-11/96, ITEP/TH-23/96, hepth/9607159.

\bibitem{M7} A.Marshakov,
{\sl Non-\-perturbative Quantum Theories and Integrable Equations},
preprint FIAN/TD-16/96, ITEP/TH-47/96, hepth/9610242.

\bibitem{MMM2}
A.Marshakov, A.Mironov and A.Morozov,
{\sl WDVV-like equations in N=2 SUSY Yang-Mills Theory},
preprint FIAN/TD-10/96, ITEP/TH-22/96, hepth/9607109, to appear in
Phys.Lett. {\bf B}.

\bibitem{MMM3}
A.Marshakov, A.Mironov and A.Morozov,
preprint FIAN/TD-15/96, ITEP/TH-46/96; hepth/9701123.

\bibitem{MMM4}
A.Marshakov, A.Mironov, A.Morozov,
preprint FIAN/TD-01/97, ITEP/TH-02/97; hepth/9701014.

\bibitem{Nekrasov}
N.Nekrasov, hepth/9609219.

\bibitem{SW1}
N.Seiberg and E.Witten, Nucl.Phys. {\bf B426} (1994) 19;
hepth/9407087.

\bibitem{SW2}
N.Seiberg and E.Witten, Nucl.Phys. {\bf B431} (1994) 484;
hepth/9408099.

\bibitem{FT} L.Faddeev and L.Takhtadjan, {\sl Hamiltonian Approach to
the Theory of Solitons}, 1986.

\bibitem{sun}
A.Klemm, W.Lerche, S.Theisen and S.Yankielowicz,
Phys.Lett. {\bf 344B} (1995) 169; hepth/9411048;\\
P.Argyres and A.Faraggi, Phys.Rev.Lett. {\bf 73}
(1995) 3931, hepth/9411057.

\bibitem{fumat}
A.Hanany and Y.Oz, hepth/9505075;\\
P.Argyres, D.Plesser and A.Shapere, hepth/9505100;\\
J.Minahan and D.Nemeschansky, hepth/9507032;\\
P.Argyres and A.Shapere, hepth/9509175;\\
A.Hanany, hepth/9509176.

\bibitem{Wils}
K.Wilson and J.Kogut, {\sl The renormalization group and
the $\epsilon $-expansion}, Phys.Rep. {\bf 12C} (1974) 75-199.

\bibitem{Krico}
I.Krichever, Func.An. \& Apps. {\bf 11} (1977) 15;\\
Uspekhi Mat.Nauk {\bf 32} (1977) 180.

\bibitem{TS}
S.Novikov, S.Manakov, L.Pitaevsky and V.Zakharov,
{\it Theory of solitons}, Moscow, Nauka 1980.

\bibitem{Dub}
B.Dubrovin, Uspekhi Mat.Nauk {\bf 36} (1981) N2, 12.

\bibitem{DKN}
B.Dubrovin, I.Krichever and S.Novikov, {\it Integrable systems - I},
{\sl Sovremennye problemy matematiki (VINITI), Dynamical systems - 4}
(1985) 179.

\bibitem{UT84}
K.Ueno and K.Takasaki, {\sl Toda lattice hierarchy},
Adv.Studies in Pure Math. {\bf 4} (1984) 1.

\bibitem{Sato}
M.Sato, RIMS Kokyuroku {\bf 439} (1981) 30;\\
M.Sato and Y.Sato, Lect.Not.Num.Appl.Anal. {\bf 5} (1982) 259;
in {\it Non-linear partial differential equations in applied
science}, Amsterdam, New York, North Holland 1983, pp 259-271.

\bibitem{SW85}
G.Segal and G.Wilson, {\sl Loop groups and equations of KdV type},
Publ.I.H.E.S. {\bf 61}(1985) 1.

\bibitem{KM1}
S.Kharchev and A.Marshakov,
in {\it String Theory, Quantum Gravity and the Unification of Fundamental
Interactions}, World Scientific (1993) 331-346.

\bibitem{KM2}
S.Kharchev and A.Marshakov,
Int. J. Mod. Phys. {\bf A10} (1995) 1219.

\bibitem{WDVV}
E.Witten, Surv.Diff.Geom. {\bf 1} (1991) 243;\\
R.Dijkgraaf, E.Verlinde and H.Verlinde, Nucl.Phys.
{\bf B352} (1991) 59.

\bibitem{KriW}
I.Krichever, Comm.Pure Appl.Math. {\bf 47} (1994) 437;
Preprint LPTENS-92-18.

\bibitem{Dubtop}
B.Dubrovin, hepth/9407018; Nucl.Phys. {\bf B379} (1992) 627.

\bibitem{Hi}
N.Hitchin, Duke.Math.Journ. {\bf 54} (1987) 91.

\bibitem{KriDu}
I.Krichever,  in the Appendix to B.Dubrovin, Uspekhi Mat.Nauk,
{\bf 36} (1981) N2, 12.

\bibitem{Skl} E.Sklyanin, J.Sov.Math. {\bf 47} (1989) 2473;
Func.Anal \& Apps. {\bf 16} (1982) 27; {\bf 17} (1983) 34.

\bibitem{KriCal}
I.Krichever, Func.Anal. \& Appl.{\bf 14} (1980) 282.

\bibitem{Rmat}
E.Sklyanin, Alg.Anal. {\bf 6} (1994) 227;\\
H.Braden and T.Suzuki, Lett.Math.Phys. {\bf 30} (1994) 147;\\
B.Enriquez and V.Rubtsov, alg-geom/9503010;\\
N.Nekrasov, hepth/9503157;\\
G.Arutyunov and P.Medvedev, hepth/9511070.

\bibitem{Ino}
V.Inozemtsev, Comm.Math.Phys. {\bf 121} (1989) 629.

\bibitem{KriUMN}
I.Krichever, Uspekhi Mat. Nauk, {\bf 44} (1989), N2, 121.

\bibitem{Kri}
I.Krichever, Func.Anal.\& Apps. {\bf 22} (1988) 37;
Comm.Math.Phys. {\bf 143} (1991) 415

\bibitem{DN}
B.Dubrovin and S.Novikov, Uspekhi Mat.Nauk, {\bf 44} (1989) N6, 29.

\bibitem{GP} A.Gurevich and L.Pitaevsky, JETP, {\bf 65} (1973)65;\\
see also S.Novikov, S.Manakov, L.Pitaevsky and V.Zakharov
''Theory of solitons", Moscow 1980.

\bibitem{Manin}
Yu.Manin, {\sl Frobenius manifolds, quantum cohomology and moduli spaces},
Preprint MPI, 1996.

\bibitem{KoMa}
M.Kontsevich and Yu.Manin, Comm.Math.Phys. {\bf 164} (1994) 525.

\bibitem{Mark}
E.Markman, Comp.Math. {\bf 93} (1994) 255;\\
R.Donagi and E.Markman, {\sl Cubics, integrable systems and Calabi-\-Yau
Threefolds}, preprint; {\sl Spectral covers, algebraically completely
integrable Hamiltonian systems and moduli of bundles}, preprint.

\bibitem{SW3}
N.Seiberg and E.Witten, hepth/9609219.

\bibitem{Ruj}
S.Ruijsenaars, {\sl Finite-dimensional Soliton Systems}, in {\it Integrable
and Super-\-Integrable Systems}, World Scientific, 1989.

\bibitem{BeMa}
A.Beilinson and Yu.Manin, Comm.Math.Phys. {\bf 107} (1986) 359.

\bibitem{Fay}
J.Fay, {\sl Theta-functions on Riemann surfaces,}
Lect. Notes Math. {\bf 352}, Springer, N.Y. 1973.

\bibitem{Mumford}
D.Mumford, {\sl Tata Lectures on Theta}, 1988.

\bibitem{Losev}
A.Losev, Theor.Math.Phys. {\bf 95} (1993) 307;\\
A.Losev and I.Polyubin, Int.J.Mod.Phys. {\bf A10} (1995) 4161-4178.

\bibitem{chr}
W.Lerche, C.Vafa and N.Warner, Nucl.Phys. {\bf B324} (1989) 427.

\bibitem{ref}
P.Griffits and J.Harris, {\sl Principles of algebraic geometry}, 1982.

\bibitem{khoze}
N.Dorey, V.Khoze and M.Mattis, hepth/9607202; hepth/9611016;\\
E.D'Hoker, I.Krichever and D.Phong, hepth/9609041.

\bibitem{KriL}
I.Krichever, hep-th/9611158.

\bibitem{strpot}
J.Harvey and G.Moore, hepth/9510182.


\end{thebibliography}
\end{document}